\documentclass[aps, pra,twocolumn,reprint,superscriptaddress]{revtex4-2}

\usepackage{graphicx}
\usepackage{dcolumn}
\usepackage{bm}
\usepackage{makecell}

\usepackage[utf8]{inputenc}
\usepackage[T1]{fontenc}
\usepackage{etoolbox}
\usepackage{subfigure}
\usepackage{amsmath,amssymb}
\usepackage{comment}
\usepackage{amssymb}

\usepackage{xcolor}
\usepackage{float}
\usepackage{amssymb}
\usepackage{hyperref}
\usepackage{comment}
\usepackage[normalem]{ulem} 
\usepackage{braket}

\begin{document}

\title{Generation of squeezed optical states via stored classical pulses in a Bose gas}

\author{Sevilay Sevin\c{c}li}
\affiliation{Department of Physics, Bilkent University, Ankara 06800, T\"{u}rkiye}
\author{Dennis R\"{a}tzel}
\affiliation{ZARM, Universit\"at Bremen, Am Fallturm 2, 28359 Bremen, Germany}
\affiliation{Vienna Center for Quantum Science and Technology, Atominstitut, TU Wien, Stadionallee 2, 1020 Vienna, Austria}
\author{Markus Krutzik}
\affiliation{Institut f\"{u}r Physik and Center for the Science of Materials Berlin (CSMB), Humboldt-Universit\"{a}t zu Berlin, Berlin, 12489, Germany}
\affiliation{Ferdinand-Braun-Institut (FBH), Gustav-Kirchoff-Str.4, 12489 Berlin, Germany}
\author{Mehmet \"{O}zg\"{u}r Oktel}
\affiliation{Department of Physics, Bilkent University, Ankara 06800, T\"{u}rkiye}
\author{Mustafa G\"{u}ndo\u{g}an}
\email[]{mustafa.guendogan@physik.hu-berlin.de}
\affiliation{Institut f\"{u}r Physik and Center for the Science of Materials Berlin (CSMB), Humboldt-Universit\"{a}t zu Berlin, Berlin, 12489, Germany}

\begin{abstract}
We propose and analyze a scheme to generate squeezed light by storing a classical probe pulse in a Bose--Einstein condensate (BEC) and exploiting the nonlinear evolution caused by atom--atom collisions during the storage time. A $\Lambda$-type optical memory interface maps a chosen temporal probe mode onto a single phase-matched collective spin wave; for a coherent input this prepares a tunable coherent spin state of a two-component BEC, with its initial spin orientation set by the stored mean excitation number and the phase relation between the probe and control fields. Collisional interactions during storage then implement one-axis-twisting dynamics and generate spin squeezing in the atomic ensemble. We account for realistic loss and finite memory and retrieval efficiencies, and model readout as a single-mode beam-splitter mapping that transfers the atomic quadrature squeezing onto a propagating optical mode. The model identifies optimal storage times and predicts that, under realistic conditions, several dB of squeezing can be transferred to the retrieved light.
\end{abstract}

\pacs{}%

\maketitle 
\section{Introduction.}
The generation of squeezed states is a cornerstone of modern quantum technologies: it enables measurement sensitivities beyond the standard quantum limit (SQL) and provides a key resource for quantum information processing. In collective spin systems, spin squeezing~\cite{Kitagawa1993,Sorensen2001,Ma2011} denotes the redistribution of quantum fluctuations such that the variance of one collective spin component is reduced below the SQL, at the expense of increased fluctuations in a non-commuting component. Spin-squeezed states have been realized in a variety of atomic platforms through both measurement-induced and interaction-driven protocols~\cite{Esteve2008,Schleier-Smith2010,Kong2020,Hosten2016b}. In quantum non-demolition (QND) schemes, dispersive probing yields information about a collective spin projection (e.g., $\hat{J}_z$) while ideally preserving its mean value; the associated measurement back-action conditionally reduces its fluctuations, and with feedback and/or cavity enhancement can produce strongly metrologically useful squeezing~\cite{Kuzmich2000,Saffman2009,Schleier-Smith2010,Hosten2016b}. Complementarily, collisional interactions in a two-component Bose--Einstein condensate (BEC) realize an effective one-axis-twisting (OAT) nonlinearity, $\hat{H}_{\mathrm{OAT}}=\hbar\chi \hat{J}_z^2$, where $\chi$ is set by the relevant scattering lengths and the trapped-mode geometry. In this unitary picture, population-dependent mean-field shifts shear the collective Bloch vector, squeezing one spin quadrature while anti-squeezing its conjugate. Interaction-driven squeezing and entanglement generation have been demonstrated in condensates using controlled collisional dynamics and internal-state interferometry~\cite{Esteve2008,Gross2010,Riedel2010,Muessel2014,Laudat2018,Berrada2013, Cassens2025}. Related spin-mixing processes in spinor condensates provide additional routes to nonclassical states, enabling twin-beam interferometry and spin-nematic squeezing~\cite{Lucke2011,Hamley2012,Kunkel2018}. These BEC-based platforms also allow spatially resolved access to many-body correlations, including Bell correlations and Einstein--Podolsky--Rosen steering~\cite{Schmied2016,Fadel2018}.

In parallel, electromagnetically induced transparency (EIT)~\cite{Boller1991,Harris1990} provides a coherent and mode-selective interface between propagating optical fields and long-lived ground-state coherences~\cite{Fleischhauer2000,Ma2017}. Implemented in many different atomic systems, EIT and related light storage techniques have enabled ultraslow and halted light as well as long-lived optical storage~\cite{Hau1999,  Fleischhauer2000, Liu2001, Gorshkov2007, Zhang2009, Katz2018, Ma2021}, offering a route to coherently map and transport collective excitations between light and matter, enabling the development of quantum memory devices~\cite{Northup2014, Lei2023} that have significant importance in quantum information science. 

In this work, we combine these two ingredients and analyze how a $\Lambda$-type EIT interface can prepare a two-component BEC in a controllable collective spin state using a classical probe pulse, while collisional interactions subsequently generate OAT-driven spin squeezing within the ensemble. Crucially, the same phase-matched atom--light coupling also defines a specific optical output mode, enabling the squeezing generated in the collective spin to be mapped onto the retrieved light field. 

\section{Physical system.}\label{sec:physical_system}
\begin{figure}[t]
  \centering
  \includegraphics[width=1\linewidth]{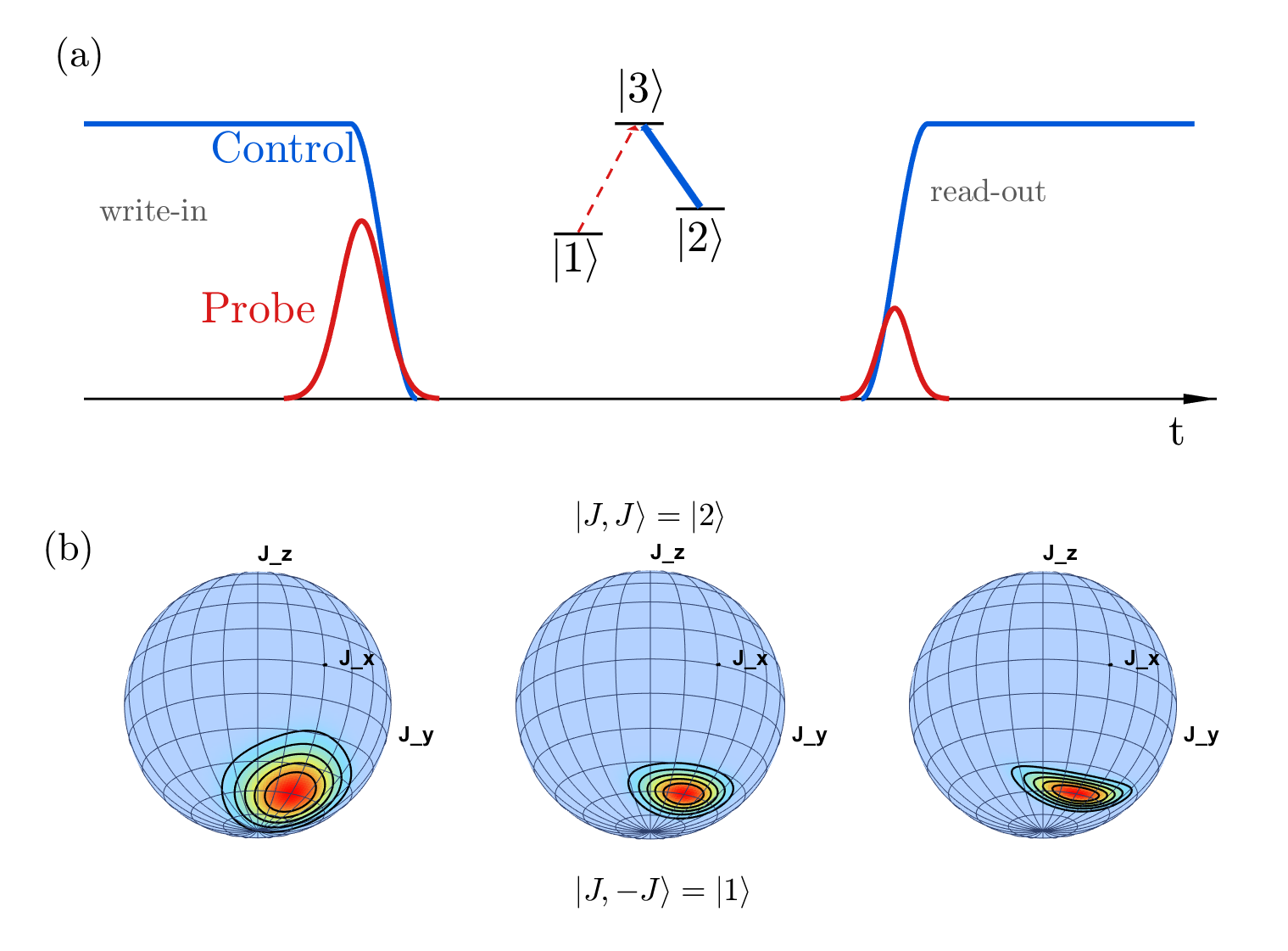}

  \caption{(a) Timing diagram of the $\Lambda$-EIT sequence. A short probe pulse (red) is incident while the control field (blue) is on (write-in). Turning the control off maps the optical excitation onto a collective $\ket{1}\!-\!\ket{2}$ ground-state coherence. After a storage interval with the light fields off, the control is turned back on (read-out) to convert the spin wave back into an emitted probe pulse. Inset: $\Lambda$-level scheme with the probe coupling $\ket{1}\!\leftrightarrow\!\ket{3}$ and the control coupling $\ket{2}\!\leftrightarrow\!\ket{3}$.
(b) Illustrative Bloch-sphere representations of the collective pseudospin formed by the two ground states $\ket{1}$ and $\ket{2}$ (south/north poles $\ket{J,-J}\equiv\ket{1}$ and $\ket{J,J}\equiv\ket{2}$). The colored patches schematically indicate the spin-noise distribution on the sphere: starting from a near-isotropic coherent-state spot and evolving into an elliptical distribution during the interaction-only evolution, illustrating the build-up of spin squeezing and the rotation of the principal noise axes.}
  \label{fig:fig1}
\end{figure}

We consider a BEC of $N_0$ identical atoms with two long-lived internal states $\{\ket{1},\ket{2}\}$, e.g. two hyperfine (or Zeeman) sublevels, forming an effective collective pseudospin $J=N_0/2$. The internal dynamics is described by the collective spin operator $\hat{\mathbf J}=\sum_{j=1}^{N_0}\hat{\mathbf j}^{(j)}$, with $\hat{\mathbf j}^{(j)}$ the spin-$1/2$ operator of atom $j$ in the $\{\ket{1},\ket{2}\}$ manifold. The state with all atoms in $\ket{1}$ corresponds to the fully polarized Dicke state $\ket{J,-J}$, i.e. the south pole of the collective Bloch sphere (Fig.~\ref{fig:fig1}b).

The two ground states $\ket{1}$ and $\ket{2}$ are coupled to a common excited state $\ket{3}$ in a $\Lambda$ configuration by a weak probe field with mean photon number of $\mu_{\mathrm{in}}$, and a strong classical control field, with corresponding optical Rabi frequencies $\Omega_p$ and $\Omega_c$, respectively. In the adiabatic EIT regime~\cite{Fleischhauer2005,Gorshkov2007} (or far-detuned Raman limit), eliminating $\ket{3}$ yields an effective interaction between one selected probe temporal mode $\hat a$ and one collective ground-state spin-wave mode $\hat S$ of the form~\cite{Choi2011coherent,Barzel2024}
\begin{equation}
\hat H_{\mathrm{int}}(t)=\hbar\,\kappa(t)\Bigl(\hat a\,\hat S^\dagger+\hat a^\dagger \hat S\Bigr),
\label{eq:Hint}
\end{equation}
where $\kappa(t)$ is controlled by the control-field envelope and detuning. For an ensemble of atoms at positions $\mathbf r_j$,
\begin{equation}
\hat S^\dagger \equiv \frac{1}{\sqrt{N_0}}\sum_{j=1}^{N_0} e^{i\Delta\mathbf k\cdot \mathbf r_j}\,
\ket{2}_j\!\bra{1},
\label{eq:Sdef}
\end{equation}
meaning that a single excitation is delocalized over $N_0$ atoms, with $\Delta\mathbf k=\mathbf k_p-\mathbf k_c$, where $\mathbf k_p$ and $\mathbf k_c$ are the probe and control wavevectors, respectively; thus, co-propagating probe/control beams ($|\Delta\mathbf k|\simeq 0$) address a spatially uniform collective mode.

Since $\kappa(t)$ is a scalar prefactor, the interaction Hamiltonian is proportional to a time-independent operator and therefore satisfies $[\hat H_{\mathrm{int}}(t),\hat H_{\mathrm{int}}(t')]=0$. The write process thus generates the beam-splitter unitary
\begin{equation}
\hat U_{\mathrm{w}}
=\exp\!\Big[-i\vartheta\big(\hat a\,\hat S^\dagger+\hat a^\dagger \hat S\big)\Big],
\end{equation}
with $\vartheta=\int_{\mathrm{write}} dt\,\kappa(t)$, which mixes the modes according to $\hat U_{\mathrm{w}}^\dagger \hat a\,\hat U_{\mathrm{w}} =\hat a\cos\vartheta-i\hat S\sin\vartheta$ and $\hat U_{\mathrm{w}}^\dagger \hat S\,\hat U_{\mathrm{w}} =\hat S\cos\vartheta-i\hat a\sin\vartheta.$ For a coherent probe input $\ket{\alpha}$ with $|\alpha|^2 = \mu_{\mathrm{in}}, $and spin-wave vacuum $\ket{0}_S$, the ideal state-level mapping is (beam splitters map coherent states to coherent states~\cite{WallsMilburn})
\begin{equation}
\hat U_{\mathrm{w}}\ket{\alpha}\ket{0}_S
=\ket{\alpha\cos\vartheta}\otimes\ket{\tilde\beta}_S,
\qquad
\tilde\beta=-i\,\alpha\sin\vartheta,
\end{equation}
so that the ideally generated spin wave is a bosonic coherent state with mean excitation $|\tilde\beta|^2$.

In practice, finite optical depth, spontaneous-emission loss, and imperfect mode matching reduce the amplitude stored in the desired collective mode. We capture these effects by defining an effective stored spin-wave coherent state $\ket{\beta}_S$ with the same phase as the ideal mapping, $\arg\beta=\arg\tilde\beta$, but reduced magnitude. Specifically, we introduce a net write-in efficiency $\eta_{\mathrm{write}}$ (defined for the chosen write sequence and thus including any incomplete transfer implicit in $\vartheta$) and a spatial mode overlap $\zeta_{\mathrm{spatial}}$, such that
\begin{equation}
\mu_{\mathrm{stored}}\equiv \langle \hat S^\dagger \hat S\rangle
=|\beta|^2
=\eta_{\mathrm{write}}\zeta_{\mathrm{spatial}}\,|\alpha|^2.
\label{eq:mueff}
\end{equation}
In the following, $\ket{\beta}_S$ is the atomic state we use as the initial condition.

\color{black}

We fix the total atom number $N_0$ in two internal modes with bosonic operators $\hat b_{1,2}$. The collective spin and number operators are $\hat J_z=\tfrac12(\hat n_2-\hat n_1)$, $\hat J_+=\hat b_2^\dagger \hat b_1$, $\hat J_-=\hat b_1^\dagger \hat b_2$ and $\hat n_1=N_0/2-\hat J_z$ and $\hat n_2=N_0/2+\hat J_z$. In the Holstein--Primakoff regime, which is equivalent to weak-probe EIT in our case, $\langle \hat S^\dagger \hat S\rangle \ll N_0$, the spin-wave creation operator can be embedded into the spin algebra as~\cite{Zhang1990}
\begin{equation}
\hat S^\dagger \simeq \frac{1}{\sqrt{N_0}}\,\hat J_+
=\frac{1}{\sqrt{N_0}}\,\hat b_2^\dagger \hat b_1.
\label{eq:HP}
\end{equation}

Within the single-mode assumption, i.e. atoms in states $\ket{1}$ and $\ket{2}$ occupy the same spatial mode, the commutator satisfies $[\hat S,\hat S^\dagger]=(\hat n_1-\hat n_2)/N_0 =1-2\hat n_2/N_0\simeq 1$ for $\langle \hat n_2\rangle\ll N_0$. which reduces to $[\hat S,\hat S^\dagger]\simeq 1$. 

\color{black}

We take the spin-wave vacuum as $\ket{0}_S\equiv\ket{N_0,0}$ (all atoms in $\ket{1}$). We describe the input probe field as an optical coherent state with complex amplitude $\alpha$, $\ket{\alpha}_{\rm in}$, normalized such that $|\alpha|^2=\mu_{\rm in}$. Assuming a phase-preserving linear write process with overall (mode-matched) storage factor $\eta_{\rm write}\zeta_{\rm spatial}$, the stored collective amplitude is $\beta=\sqrt{\eta_{\rm write}\zeta_{\rm spatial}}\,\alpha$, so that the mean stored excitation number is $\mu_{\rm stored}=|\beta|^2=\eta_{\rm write}\zeta_{\rm spatial}\mu_{\rm in}$.

A stored spin-wave coherent state is obtained by displacing the spin-wave  vacuum, $\ket{\beta}_S\equiv \hat D_S(\beta)\ket{0}_S$, with the standard displacement operator $\hat D_S(\beta)=\exp(\beta \hat S^\dagger-\beta^*\hat S)$. Using the normally ordered identity $\hat D_S(\beta)=e^{-|\beta|^2/2}e^{\beta \hat S^\dagger}e^{-\beta^*\hat S}$ and the vacuum property $\hat S\ket{0}_S=0$, the rightmost factor acts trivially on $\ket{0}_S$, which is why the $-\beta^*\hat S$ term does not appear explicitly once the operator is applied to the vacuum. In the Holstein--Primakoff (weak-excitation) regime $\mu_{\rm stored}=|\beta|^2\ll N_0$, we further use $\hat S^\dagger\simeq(\hat b_2^\dagger\hat b_1)/\sqrt{N_0}$, yielding the fixed-$N_0$ ladder form
\begin{align}
\ket{\beta}_S
&=\hat D_S(\beta)\ket{0}_S
= e^{-|\beta|^2/2}\exp(\beta \hat S^\dagger)\ket{0}_S
\nonumber\\
&\simeq e^{-|\beta|^2/2}\exp\!\Big(\tfrac{\beta}{\sqrt{N_0}}\,\hat b_2^\dagger \hat b_1\Big)\ket{N_0,0}.
\label{eq:beta_to_SU2}
\end{align}

Atomic coherent states, or coherent spin states (CSS), may be defined as rotations $\ket{\theta,\varphi_0}=e^{-i\varphi_0\hat J_z}e^{-i\theta\hat J_y}\ket{J,-J}$ and admit an equivalent ladder form~\cite{Zhang1990} $(1+|\tau|^2)^{-J}\exp(\tau\hat J_+)\ket{J,-J}$ with $\tau=e^{i\varphi_0}\tan(\theta/2)$; the derivation and conventions are summarized in Appendix~\ref{app:alpha_mapping}. Comparison with Eq.~\eqref{eq:beta_to_SU2} identifies $\tau=\beta/\sqrt{N_0}=\sqrt{\eta_{\rm write}\zeta_{\rm spatial}/N_0}\,\alpha$ and $\varphi_0=\arg\beta$ (set by the probe/control phase reference). In the same weak-excitation regime $|\tau|^2\ll 1$, the CSS normalization satisfies $(1+|\tau|^2)^{-J}\simeq\exp(-J|\tau|^2)=\exp(-|\beta|^2/2)$, which matches the bosonic prefactor in $\ket{\beta}_S$ (Appendix~\ref{app:alpha_mapping}). For a coherent spin state, the mean transferred population is $\mu_{\rm stored}=\langle \hat n_2\rangle=N_0\sin^2(\theta/2)\simeq N_0(\theta/2)^2$ for $\mu_{\rm stored}\ll N_0$, hence $\theta\simeq 2\sqrt{\mu_{\rm stored}/N_0}=2\sqrt{\eta_{\rm write}\zeta_{\rm spatial}\mu_{\rm in}/N_0}$. Thus, the mean input excitation $\mu_{\rm in}$ together with $\eta_{\rm write}$ and $\zeta_{\rm spatial}$ sets the initial, small tilt $\theta$, while the probe/control phase sets the azimuth $\varphi_0$.

During the storage stage the optical fields are off and the dynamics of the two-component condensate is governed by single-particle energies and $s$-wave collisional interactions. In a two-mode description we introduce bosonic mode operators $\hat b_i$ for the two internal states ($i=1,2$) and number operators $\hat n_i=\hat b_i^\dagger\hat b_i$. The effective Hamiltonian reads
\begin{equation}
\hat H=\hat H_{0}+\hat H_{\rm int},
\end{equation}
with the single-particle part
\begin{equation}
\hat H_{0}\simeq E_\phi(\hat n_1+\hat n_2)+\epsilon_1\hat n_1+\epsilon_2\hat n_2,
\label{eq:H0_twomode_rewrite_full}
\end{equation}
and the collisional interaction
\begin{equation}
\hat H_{\rm int}
=\frac{U_{11}}{2}\hat b_1^{\dagger 2}\hat b_1^{2}
+\frac{U_{22}}{2}\hat b_2^{\dagger 2}\hat b_2^{2}
+U_{12}\hat b_1^\dagger \hat b_2^\dagger \hat b_1 \hat b_2,
\label{eq:Hcoll_twomode_rewrite_full}
\end{equation}
where $E_\phi$ is the common single-particle mode energy (kinetic plus trap energy) of the shared spatial mode, $\epsilon_i$ are the internal-state energy offsets in the rotating frame (only the difference $\epsilon_1-\epsilon_2$ affects the relative phase evolution), and $U_{ij}$ are the effective interaction matrix elements (e.g. $U_{ij}=g_{ij}\!\int d^3r\,|\phi(\mathbf r)|^4$ in a single-spatial-mode picture with $g_{ij}=4\pi\hbar^2 a_{ij}/m$, where $a_{ij}$ is the S-wave scattering length between the states $\ket{1}$ and $\ket{2}$.

For fixed total atom number $\hat n_1+\hat n_2=N_0$, the term $E_\phi(\hat n_1+\hat n_2)=E_\phi N_0$ is an additive constant that contributes only a global phase and will be dropped. Substituting $\hat J_z$ and $\hat n_{1,2}$ into $\hat H_0+\hat H_{\rm int}$ (discarding additive constants) yields the OAT Hamiltonian
\begin{equation}\label{eq:H_OAT}
\hat H_{\rm OAT}=\hbar\chi\,\hat J_z^2+\hbar\Omega\,\hat J_z+\mathrm{const.},
\end{equation}
with $\chi=(U_{11}+U_{22}-2U_{12})/(2\hbar)$ and $\Omega=(\epsilon_2-\epsilon_1+(N_0-1)(U_{22}-U_{11})/2)/\hbar$. The nonlinear term $\propto \hat J_z^2$ is the one-axis-twisting interaction that generates spin squeezing, while the linear term $\propto \hat J_z$ corresponds to a collective rotation about the $z$ axis (differential phase accumulation) and can be removed by working in a rotating frame with respect to $\hbar\Omega\hat J_z$ or by appropriate phase referencing.

\section{Squeezing dynamics, loss, and optical readout}
\label{sec:squeezing_dynamics}

\color{black}
After the write process prepares an initial coherent spin state $\ket{\theta,\varphi_0}$ (Sec.~II), the two-component condensate evolves under the effective OAT Hamiltonian, Eq.~\eqref{eq:H_OAT}. We work throughout in the rotating frame of the linear term $\hbar\Omega \hat J_z$ (equivalently, we absorb this deterministic $z$ rotation into the phase reference) and thus retain only the nonlinear term in Eq.~\ref{eq:H_OAT}. The interaction-induced phase shift then depends on the population difference, which shears the collective noise distribution on the Bloch sphere and generates squeezing in an optimally chosen quadrature. Because $[\hat H_{\rm OAT},\hat J_z]=0$, unitary OAT conserves $\hat J_z$ and hence the instantaneous populations; the nontrivial dynamics is therefore encoded in the transverse spin components and their correlations with $\hat J_z$. As elaborated in Sec.~\ref{sec:physical_system}, we focus on the weak-excitation regime $\mu_{\rm stored}\ll N_0$, where the initial tilt is small and the relevant squeezing dynamics is well captured by transverse fluctuations around the mean spin.
\color{black}

\subsection{Exact moments from Heisenberg correlators}
\label{sec:exact_moments}

To quantify squeezing we need the first moments $\langle\hat{\mathbf J}\rangle$ and the symmetrized second moments that form the covariance matrix (Sec.~\ref{sec:wineland_singlecol}). We work directly in the fixed-$J$ Dicke manifold and compute these moments exactly without propagating the full many-body state. For one-axis twisting (OAT) $\hat J_z$ is conserved and the ladder operators acquire phases depending only on the $\hat J_z$ eigenvalue $m$. This reduces the problem to evaluating a small set of Heisenberg-picture correlators as finite Dicke-basis sums.

We introduce the following four expectation values: 
\begin{align}
E_1(t) &\equiv \langle \hat J_+(t)\rangle, \nonumber\\
E_2(t) &\equiv \langle \hat J_-(t)\rangle, \nonumber\\
E_3(t) &\equiv \langle \hat J_+^2(t)\rangle, \nonumber\\
E_4(t) &\equiv \langle \hat J_-^2(t)\rangle .
\label{eq:E1toE4_defs}
\end{align}
They determine the transverse mean spin and the second moments within the $x$--$y$ plane. Mixed moments that involve $\hat J_z$ are expressed through additional number-weighted correlators $E_5,\ldots,E_{12}$. Their explicit forms are collected in Appendix~\ref{app:correlators_defs}. From $E_1$ and $E_2$ we obtain the transverse mean spin components
\begin{align}
\langle \hat J_x\rangle &= \tfrac12\big(E_1+E_2\big), \nonumber\\
\langle \hat J_y\rangle &= \tfrac{1}{2i}\big(E_1-E_2\big).
\label{eq:Jxy_from_E12}
\end{align}
For the transverse second moments we use the fixed-$J$ relations $\hat J_+\hat J_-=\hat J^2-\hat J_z^2+\hat J_z$ and $\hat J_-\hat J_+=\hat J^2-\hat J_z^2-\hat J_z$. This gives
\begin{align}
\langle \hat J_x^2\rangle
&=
\frac{E_4+E_3}{4}
+\frac12\Big(J(J+1)-\langle \hat J_z^2\rangle\Big),
\label{eq:Jx2_from_E}
\\
\langle \hat J_y^2\rangle
&=
-\frac{E_4+E_3}{4}
+\frac12\Big(J(J+1)-\langle \hat J_z^2\rangle\Big),
\label{eq:Jy2_from_E}
\\
\left\langle \tfrac12\{\hat J_x,\hat J_y\}\right\rangle
&=
\frac{E_3-E_4}{4i}.
\label{eq:JxJy_from_E}
\end{align}
The two-quantum coherences $E_3$ and $E_4$ set the anisotropy and orientation of the transverse noise ellipse. The remaining term $J(J+1)-\langle \hat J_z^2\rangle$ provides the isotropic offset fixed by the spin length and the $\hat J_z$ statistics.

To determine the orientation of the squeezing ellipse (i.e. the principal axes of the transverse fluctuations) with respect to the mean-spin direction, we also require the mixed symmetrized moments  
with $\hat J_z$. Using the number-weighted correlators from Appendix~\ref{app:correlators} we define
\begin{align}
\mathcal C_{xz}(t)
&\equiv (E_{12}-E_{11})+(E_{10}-E_{9})
\nonumber\\
&\phantom{\equiv}+(E_{7}-E_{5})+(E_{8}-E_{6}),
\label{eq:Cxz_def}
\\
\mathcal C_{yz}(t)
&\equiv (E_{11}-E_{12})+(E_{10}-E_{9}) \nonumber\\
&\phantom{\equiv}+(E_{7}-E_{5})+(E_{6}-E_{8}).
\label{eq:Cyz_def}
\end{align}
The corresponding anticommutators are 
\begin{align}
\left\langle \tfrac12\{\hat J_x,\hat J_z\}\right\rangle
&= \frac{\mathcal C_{xz}(t)}{8},
\label{eq:JxJz_from_E}
\\
\left\langle \tfrac12\{\hat J_y,\hat J_z\}\right\rangle
&= \frac{\mathcal C_{yz}(t)}{8i}.
\label{eq:JyJz_from_E}
\end{align}

Together with the longitudinal moments $\langle \hat J_z\rangle$ and $\langle \hat J_z^2\rangle$, Eqs.~\eqref{eq:Jxy_from_E12}--\eqref{eq:JyJz_from_E} fully specify the covariance matrix $\Gamma(t)$ and hence the minimal transverse variance $\lambda_{\min}(t)$ (Sec.~\ref{sec:wineland_singlecol}). In the following subsections we extend this unitary description to include particle loss and to map the resulting atomic squeezing onto the retrieved optical field.

\subsection{Covariance matrix and transverse squeezing}
\label{sec:wineland_singlecol}

Quantum fluctuations of the collective spin are summarized by the symmetrized covariance matrix
\begin{align}
\Gamma_{ij}(t)
&\equiv
\frac12\Big\langle
\Delta\hat J_i\,\Delta\hat J_j
+\Delta\hat J_j\,\Delta\hat J_i
\Big\rangle,
\label{eq:cov_def_singlecol}
\end{align}
where, $\Delta\hat J_i=\hat J_i-\langle\hat J_i\rangle$, with $i,j\in\{x,y,z\}$. For any real unit vector $\mathbf u$, the variance of the spin projection $\hat J_{\mathbf u}\equiv \mathbf u\cdot \hat{\mathbf J}$ is given by $\mathrm{Var}(\hat J_{\mathbf u})=\mathbf u^{\top}\Gamma\,\mathbf u$.

The metrological signal is carried by the mean spin vector $\mathbf J(t)=\langle \hat{\mathbf J}(t)\rangle$. Noise relevant for phase estimation is the noise in directions transverse to $\mathbf J(t)$, i.e. in the plane orthogonal to the unit vector $\mathbf n(t)=\mathbf J(t)/|\mathbf J(t)|$. To isolate these fluctuations we introduce the projector $\mathbf P(t)=\mathbb I_3-\mathbf n(t)\mathbf n(t)^{\top}$ and define the projected covariance $\Gamma_\perp(t)=\mathbf P(t)\,\Gamma(t)\,\mathbf P(t)$. This projection removes the component parallel to $\mathbf n(t)$ and retains fluctuations within the transverse plane. Accordingly, $\Gamma_\perp(t)$ has one null direction along $\mathbf n(t)$ and two nonzero eigenvalues, which are the variances along the principal axes of the transverse noise ellipse. We denote by $\lambda_{\min}(t)$ the smaller of these two eigenvalues. Equivalently, $\lambda_{\min}(t)$ is the minimum variance of a transverse spin component $\hat J_{\mathbf u}=\mathbf u\cdot\hat{\mathbf J}$ over all unit vectors $\mathbf u$ satisfying $\mathbf u\cdot\mathbf n(t)=0$.

We characterize atomic squeezing by the smallest fluctuation of a collective-spin component orthogonal to the instantaneous mean spin $\mathbf J(t)\equiv\langle\hat{\mathbf J}(t)\rangle$~\cite{Kitagawa1993}. Let $\lambda_{\min}(t)$ denote the smaller eigenvalue of the covariance matrix projected onto the plane perpendicular to $\mathbf J(t)$, i.e.\ the minimal transverse spin variance $\Delta J_{\perp,\min}^2(t)$. For a coherent spin state with the same mean-spin length, the transverse variance is $|\mathbf J(t)|/2$. We therefore define the dimensionless squeezing parameter
\begin{equation}
v_{A,\min}(t)\equiv \frac{\lambda_{\min}(t)}{|\mathbf J(t)|/2},
\label{eq:vAmin_def}
\end{equation}
so that $v_{A,\min}=1$ corresponds to the CSS, i.e. shot-noise, level. In the unitary, high-contrast limit $|\mathbf J(t)|\simeq J=N/2$, this reduces to the Kitagawa--Ueda definition $\xi_S^2=4\,\Delta J_{\perp,\min}^2/N$~\cite{Kitagawa1993}.This is the quantity entering our optical readout model (details in Sec.~\ref{sec:readout_singlecol}): when the local-oscillator phase is aligned with the minimal-noise axis, the retrieved optical quadrature variance is obtained by mixing $v_{A,\min}(t)$ with vacuum according to the readout efficiency.

\subsection{Population loss and injected transverse noise}
\label{sec:loss_singlecol}

We include irreversible losses at the level of mean populations by evolving $N_i(t)=\langle \hat n_i\rangle$ with rate equations that include one-, two-, and three-body processes. These deterministic equations set the longitudinal first moment $\langle \hat J_z\rangle=[N_2(t)-N_1(t)]/2$ and define an instantaneous effective spin length $J_{\mathrm{stored}}(t)=N_e(t)/2$ with $N_e(t)=N_1(t)+N_2(t)$.  We model the population dynamics as

\begin{equation}
\label{eq:loss_odes_singlecol}
\dot N_i(t)
= -K_{1,i}\,N_i
-\sum_j K_{2,ij}\,N_i N_j
-\sum_{j,k} K_{3,ijk}\,N_i N_j N_k,
\end{equation}
where $K_{1,i}$ are one-body loss rates (units of s$^{-1}$), $K_{2,ij}$ are two-body loss coefficients (units of m$^{3}$/s), and $K_{3,ijk}$ are three-body loss coefficients (units of m$^{6}$/s). The indices indicate the internal states participating in the event. See Appendix~\ref{app:loss_rescaling} for detailed explanation and numerical values used in our simulations. 

\begin{figure}[t]
  \centering

  \subfigure[]{
    \includegraphics[width=0.85\linewidth]{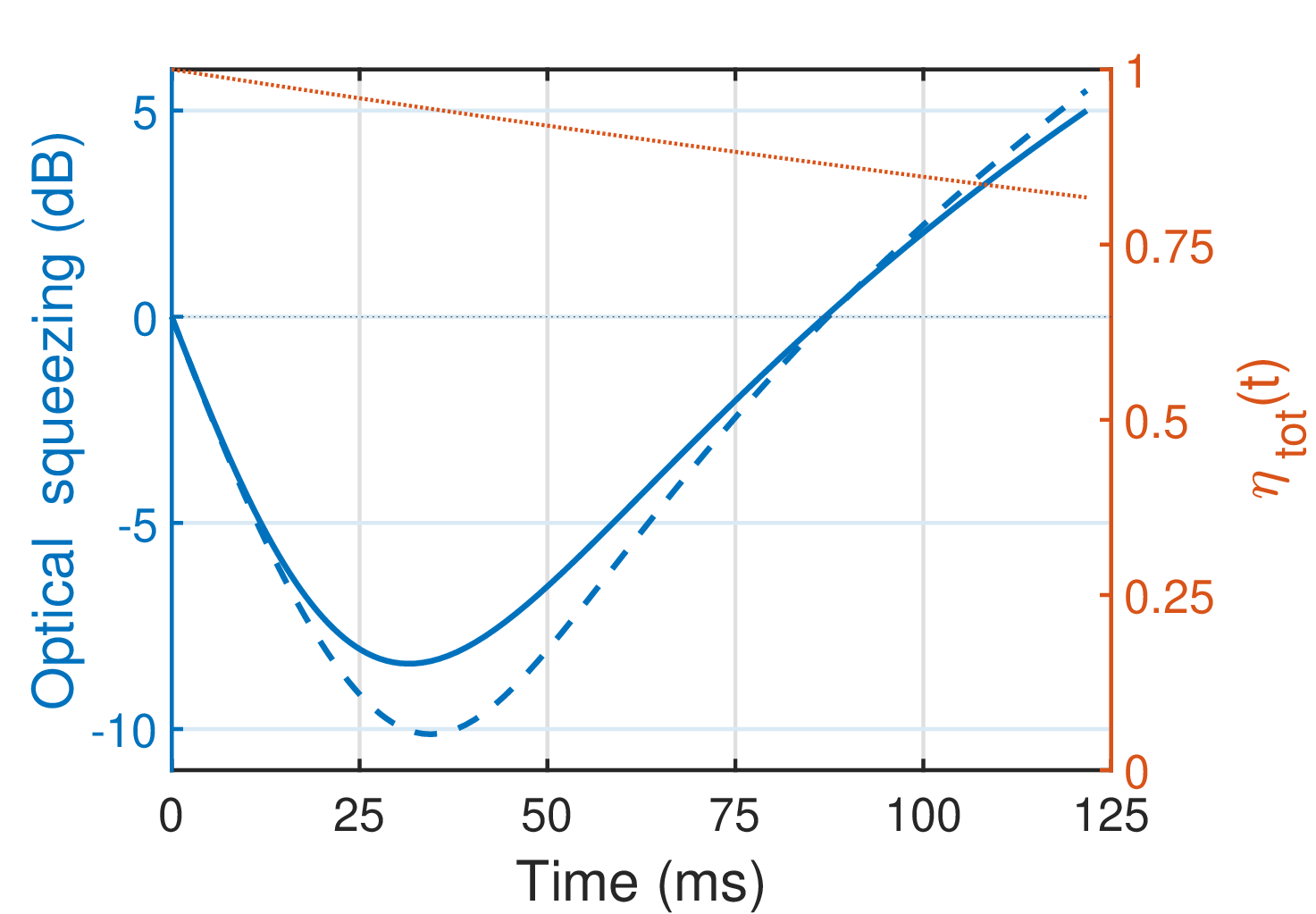}
    \label{fig:top2}
  }

  \vspace{0.6em}

  \subfigure[]{
    \includegraphics[width=0.85\linewidth]{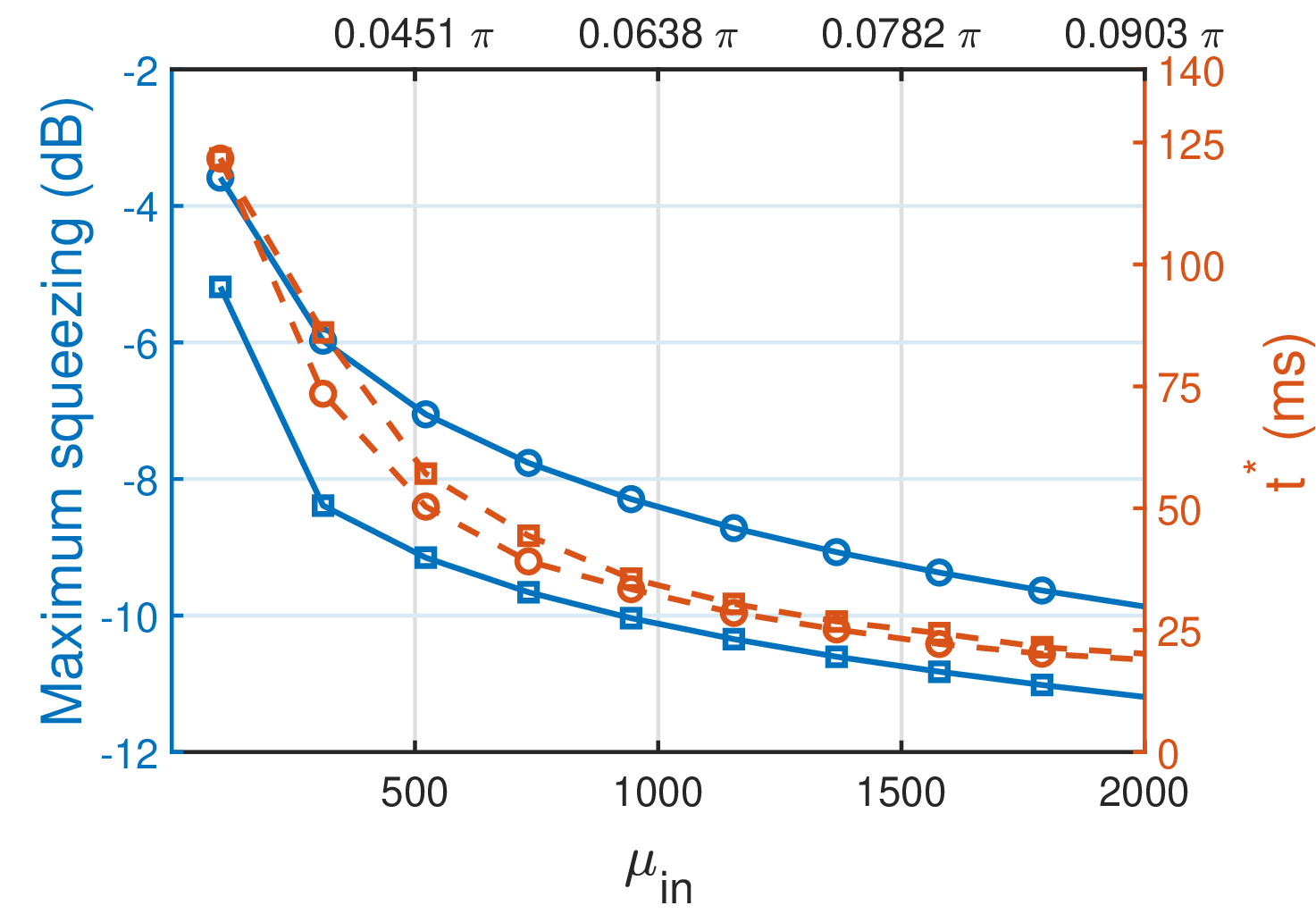}
    \label{fig:bottom2}
  }
\caption{(a) Time evolution of the retrieved optical variance $10\log_{10}V_{\mathrm{opt}}(t)$ (blue, left axis) for $\mu_{\mathrm{in}}=10^3$; solid (lossy) and dashed (lossless) curves are shown. The dotted red curve (right axis) shows the total memory efficiency $\eta_{\mathrm{tot}}(t)$ for the lossy case. (b) Best (minimum) retrieved squeezing $\min_t[10\log_{10}V_{\mathrm{opt}}(t)]$ (blue, left axis) and the corresponding optimal time $t^*$ (red, right axis) versus $\mu_{\mathrm{in}}$; the upper axis indicates the equivalent initial spin-tilt angle $\theta$, corresponding to the respective $\mu_{\mathrm{in}}$. Circles: lossy case, squares: lossles case. }
  \label{fig:fig2}
\end{figure}

Loss also reduces collective $\ket{1}\!-\!\ket{2}$ coherence. We model this by a minimal, number-consistent renormalization of the unitary OAT correlators $E_k^{(u)}(t)$ obtained from unitary evolution under $\hat H_{\rm OAT}$ (i.e., with all loss channels set to zero) on the fixed-$J$ manifold (Appendix~\ref{app:loss_rescaling}). Concretely, we write $E_k(t)=S_k(t)\,E_k^{(u)}(t)$, where the prefactor $S_k(t)$ is constructed from the instantaneous mean populations $N_{1,2}(t)$ and depends only on whether the correlator contains one or two ladder operators $\hat J_\pm$ and whether it is weighted by $\hat n_1$ or $\hat n_2$. The explicit mapping $k\mapsto S_k(t)$ is given in Appendix~\ref{app:loss_rescaling}. A useful by-product of this construction is a simple, ``memory-efficiency'' factor. The stored spin wave is proportional to the collective coherence operator $\hat J_+\equiv \hat b_2^\dagger \hat b_1$. For a fully phase-coherent two-component state with populations $N_1$ and $N_2$, the collective coherence reaches $|\langle \hat J_+\rangle|=\sqrt{N_1N_2}$ \cite{Arecchi1972}. Since the retrieved probe field amplitude in EIT readout is linear in the stored Raman coherence \cite{Fleischhauer2000,Dutton2004}, the retrieved intensity (and hence an effective quadrature-mapping strength) scales as $|\langle \hat J_+\rangle|^2\propto N_1N_2$. This motivates the intrinsic, coherence-limited factor
\begin{equation}
\eta_{\rm coh}(t)=\frac{N_1(t)N_2(t)}{N_{1,0}N_{2,0}}.
\label{eq:eta_coh}
\end{equation}
In other words, the same population-loss model that determines $N_{1,2}(t)$ also determines the natural ``memory decay'' through loss of coherence, $\eta_{\rm coh}(t)$. We will use this quantity as time-dependent memory efficiency, $\eta_{\rm tot}(t) = \eta_{\rm coh}(t)\times \eta_{\rm write}\times\eta_{\rm read}$, as shown by the orange curve in Fig.~\ref{fig:fig2}(a).

All first and second moments are then reconstructed from Eqs.~\eqref{eq:Jxy_from_E12}--\eqref{eq:JyJz_from_E}, using the rescaled correlators $E_k(t)$ and replacing $J\to J_{\mathrm{stored}}(t)$ in the static $J(J+1)$ contributions. Importantly, because the coherence reduction enters already at the level of the rescaled correlators (through $R_1$ and $R_2$), the resulting $v_{A,\min}(t)$ refers directly to the retrievable collective mode at time $t$; we therefore do not apply an additional $\eta_{\rm coh}(t)$ factor again at the optical beam-splitter stage (Sec.~\ref{sec:readout_singlecol}).

The rescaling above accounts for average damping of mean values but not for the extra fluctuations generated by stochastic loss events (quantum jumps). Following Refs.~\cite{Li2008,Li2009}, we include their leading-order effect as an approximately isotropic diffusion in the plane transverse to the instantaneous mean spin. We implement this by the covariance update
\begin{align}
\boldsymbol{\Gamma}(t+\mathrm dt)
=
\boldsymbol{\Gamma}(t)
+
\frac{|\langle \hat{\mathbf J}(t)\rangle|^2}{N_e(t)}\,\frac{s_q(t)}{3}\,\mathrm dt\,\mathbf P(t),
\label{eq:LCS_noise_singlecol}
\end{align}
where $s_q(t)=\gamma_1(t)+2\gamma_2(t)+3\gamma_3(t)$ is the weighted sum of effective one-, two-, and three-body loss rates evaluated on the instantaneous mean state (explicit forms of $\gamma_i$ are given in Appendix~\ref{app:loss_rescaling}). In our numerics, $\boldsymbol{\Gamma}(t)$ in Eq.~\eqref{eq:LCS_noise_singlecol} denotes only the accumulated loss-induced noise contribution to the covariance (initialized to zero); it is added to the covariance reconstructed from the rescaled OAT correlators to form the total covariance used to compute $\lambda_{\min}(t)$ and hence $v_{A,\min}(t)$. This makes explicit that the rescaling captures average coherence damping, whereas Eq.~\eqref{eq:LCS_noise_singlecol} accounts for the additional transverse fluctuations injected by random loss events.

\

\subsection{Optical readout: mapping atomic to optical squeezing}
\label{sec:readout_singlecol}

The read process is the time-reverse of the write-in beam splitter and maps the collective Raman coherence (spin-wave mode) back onto a single optical output mode. In the adiabatic EIT picture, this corresponds to rotating the dark-state polariton from matter-like to light-like by turning on the control field, thereby coherently transferring the stored atomic excitation back to the probe field mode~\cite{Fleischhauer2000,Dutton2004}. We therefore model the external readout and detection as a linear loss channel of efficiency $\eta_{\rm read}$ acting on the relevant atomic quadrature, with the unused port fed by vacuum.

The retrieval of non-classical optical states from atomic ensembles has been demonstrated in a number of experiments, including the storage and recall of squeezed and entangled light fields (and related non-classical photonic inputs)~\cite{Appel2008, Honda2008, Choi2008, Lettner2011, Seri2017}. In those works, however, the non-classicality is primarily injected into the memory via the input light and then preserved during storage. In contrast, our proposal uses the memory interface as part of the state-preparation mechanism: the write--store--read sequence prepares a two-component BEC whose intrinsic interactions generate spin squeezing during storage, which is subsequently mapped onto the retrieved optical mode. 

When the local-oscillator (LO) phase is aligned with the instantaneous minimal-noise axis of the atomic state, the detected optical quadrature variance is
\begin{equation}
V_{\rm opt}(t)=\bigl[1-\eta_{\rm read}\bigr]+\eta_{\rm read}\,v_{A,\min}(t).
\label{eq:Vopt_singlecol}
\end{equation}
Equation~\eqref{eq:Vopt_singlecol} is the usual beam-splitter relation for variances: a retrieval channel of efficiency $\eta_{\rm read}$ transmits the squeezed atomic quadrature with weight $\eta_{\rm read}$ and admixes vacuum with weight $1-\eta_{\rm read}$, thereby reducing the observable squeezing. In our model, the atomic variance $v_{A,\min}(t)$ already includes the time-dependent in-medium degradation of the spin wave due to loss (including the coherence-limited reduction absorbed into the correlator rescaling in Sec.~\ref{sec:loss_singlecol}) and any write-in imperfections; the conversion to the detected optical mode is therefore described solely by the additional external efficiency $\eta_{\rm read}$.

\section{Results}
\label{sec:results_singlecol}

To benchmark our model with experimentally grounded parameters, we simulate a two-component $^{23}\mathrm{Na}$ BEC in an optical dipole trap. We adopt the $\Lambda$-scheme EIT and interaction parameters at the working bias field from the long-lived EIT memory experiment of Ref.~\cite{Zhang2009}, but use a more compact, symmetrical cloud  (Thomas--Fermi diameter $d_{\mathrm{TF}}\simeq 10~\mu\mathrm{m}$) with atom number ($N_0=1\times 10^{5}$) and size comparable to the $^{87}$Rb BEC optical memory experiment~\cite{Saglamyurek2021}. All results in this section are obtained from numerical simulations of the effective two-component collective-spin model introduced above: we evaluate the unitary OAT dynamics and combine it with numerical integration of the mean population-loss equations and the associated injected transverse noise.

We model a $\Lambda$-type EIT interface on the sodium D$_1$ line, $3S_{1/2}\!\rightarrow 3P_{1/2}$, with ground states $\ket{1}\equiv\ket{3S_{1/2},F=1,M=0}$ and $\ket{2}\equiv\ket{3S_{1/2},F=2,M=-2}$, coupled via the excited state $\ket{3}\equiv\ket{3P_{1/2},F'=1,M'=-1}$; the coupling field is resonant with $\ket{2}\!\rightarrow\!\ket{3}$ and the probe field addresses $\ket{1}\!\rightarrow\!\ket{3}$~\cite{Zhang2009}. The subsequent internal-state evolution is modeled as OAT with $\chi$ computed from the elastic $s$-wave scattering lengths at the working bias field $B\simeq 132.4~\mathrm{G}$, namely $a_{11}=2.8~\mathrm{nm}$, $a_{22}=3.4~\mathrm{nm}$, and $a_{12}=3.4~\mathrm{nm}$~\cite{Zhang2009}. At this field, inelastic losses in the $\ket{1}$--$\ket{2}$ channel are strongly suppressed: $\mathrm{Im}(a_{12})$ approaches zero and it is assumed to be $-1\times10^{-3}\times a_0$ in our simulations, and $\mathrm{Im}(a_{11})$ and $\mathrm{Im}(a_{22})$ are negligible~\cite{Zhang2009}. Numerical values of $K$ parameters of Eq.~\ref{eq:loss_odes_singlecol} are detailed in Appendix~\ref{app:loss_rescaling}. At these elastic parameters one has $a_{12}^2>a_{11}a_{22}$, i.e.\ the mixture is formally immiscible, consistent with the phase-separation dynamics reported in Ref.~\cite{Zhang2009} for strong stored-pulse imprints. In contrast, our simulations focus on the weak-excitation (Holstein--Primakoff) regime $n_2\ll n_1$, for which the spin-demixing instability rate is parametrically suppressed by the minority density $n_2$ and, for our parameters, the most unstable demixing wavelength exceeds the cloud size; see Appendix~\ref{app:demixing} for a quantitative estimate based on the Bogoliubov spectrum of a binary condensate~\cite{Kasamatsu2004MI,Timmermans1998PhaseSeparation}. This separation-of-timescales requirement is essential for spin squeezing: once significant demixing occurs, the spatial overlap $\int d^3r\,|\phi_1(\mathbf r)|^2|\phi_2(\mathbf r)|^2$ drops rapidly, which quenches the effective intercomponent nonlinearity and coherence on which OAT squeezing relies; thus we require the demixing time $\tau_{\mathrm{MI}}$ to exceed the squeezing-development time $t_{\mathrm{sq}}$ (approximately the time to reach the minimum of $v_{A,\min}(t)$) so that appreciable squeezing can build up before overlap is lost. Optical readout is modeled by a constant read-out efficiency $\eta_{\mathrm{read}}$, and unless stated otherwise we set $\zeta_{\mathrm{spatial}}=\eta_{\mathrm{write}}=\eta_{\mathrm{read}}=1$.

\begin{figure}[t]
  \centering
  \includegraphics[width=1\linewidth]{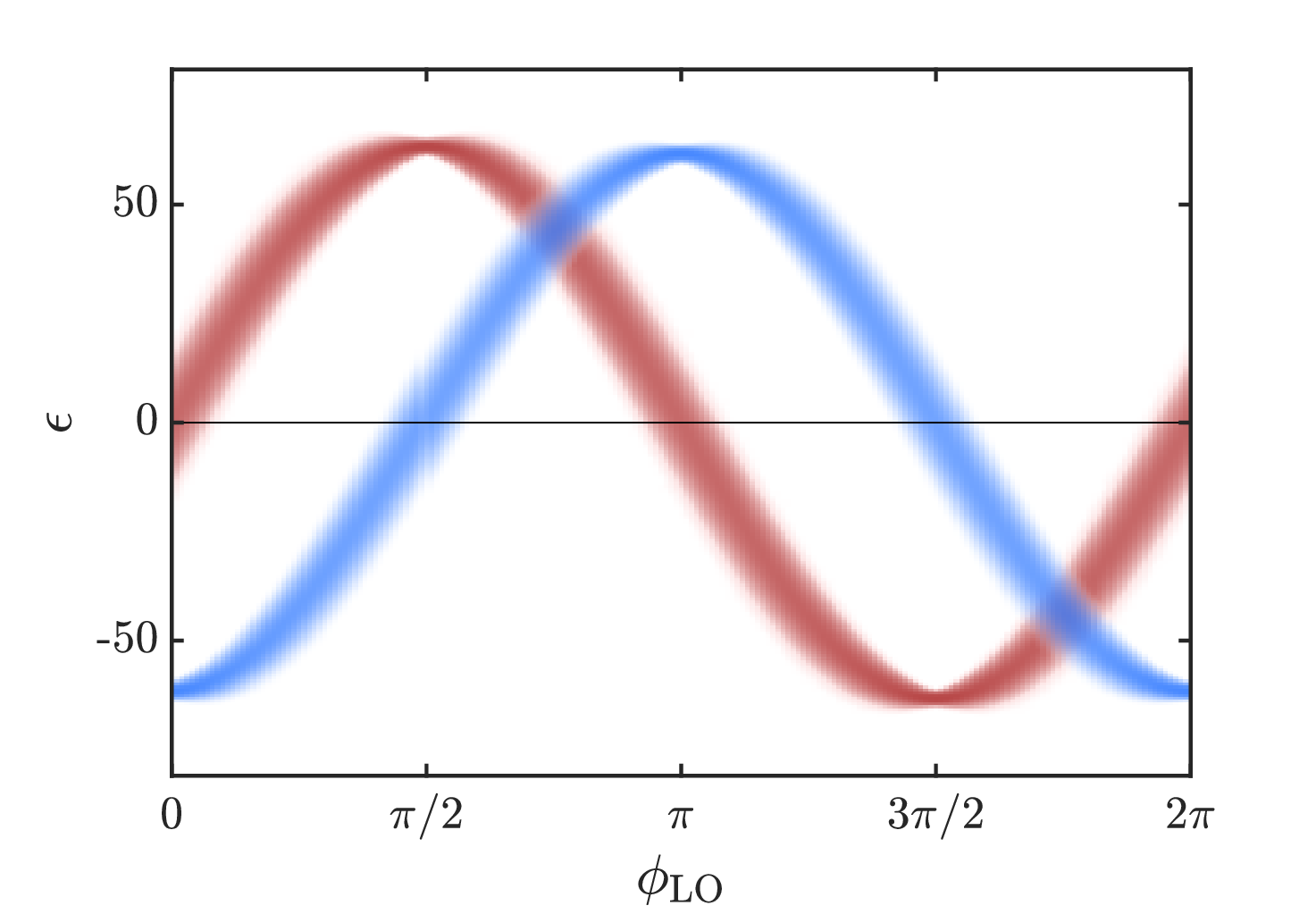}
  \caption{Simulated balanced-homodyne LO-phase scan of the retrieved optical mode at the optimal storage time $t^\ast$. For each $\phi_{\mathrm{LO}}$ we generate $N_{\mathrm{sample}}$ single-shot quadrature outcomes $\epsilon$ by sampling $\epsilon=\mu(\phi_{\mathrm{LO}})+\sqrt{V_{\mathrm{det}}(\phi_{\mathrm{LO}})}\,\xi$ with $\xi\sim\mathcal{N}(0,1)$, where $\mu(\phi_{\mathrm{LO}})=A_{\mathrm{coh}}\cos(\phi_{\mathrm{LO}}-\phi_{\mathrm{coh}})$ and $V_{\mathrm{det}}(\phi_{\mathrm{LO}})=(1-\eta_{\mathrm{read}})+\eta_{\mathrm{read}}\,v_A(\phi_{\mathrm{LO}})$. The heatmaps show $\log_{10}(\mathrm{counts}+1)$ from binned samples: red denotes the lossless reference ($\mathrm{Im}(a_{12})=0$, all loss channels off) and blue (shifted by $\pi/2$ to avoid overlap with the red curve) denotes the lossy $^{23}$Na case (loss channels on; representative $\mathrm{Im}(a_{12})=-0.001\,a_0$ used to set the inter-state two-body loss in the model).}
  \label{fig:LO_scan}
\end{figure}

Figure~\ref{fig:fig2}(a) summarizes the central prediction of our model: an optical pulse stored by EIT in a BEC prepares a CSS that subsequently develops interaction-drivenspin squeezing,  and this atomic squeezing can be mapped back onto a single, well-defined optical output mode upon readout. We quantify squeezing by the minimal transverse atomic variance $v_{A,\min}(t)$ (normalized to the CSS noise, and shown in dB as $10\log_{10}v_{A,\min}$), and convert it to the detected optical variance via Eq.~\eqref{eq:Vopt_singlecol}, $V_{\mathrm{det},\min}(t)=(1-\eta_{\mathrm{read}})+\eta_{\mathrm{read}},v_{A,\min}(t)$. The dashed blue curve shows a lossless reference evolution with $\mathrm{Im}(a_{12})=0$ and all inelastic channels disabled ($K_{1}=K_{2}=K_{3}=0$), while the solid blue curve includes inelastic processes and enabling one-, two-, and three-body loss channels (see Appendix~\ref{app:loss_rescaling} for details and sources). For an input pulse with mean photon number $\mu_{\mathrm{in}}=10^3$ (corresponding to an initial tilt $\theta\simeq0.064\pi$ for the parameters used here), the lossless dynamics reaches a minimum of $\simeq-10.1$~dB at $t\simeq 35~\mathrm{ms}$ before over-twisting causes the variance to rise, whereas including loss both injects additional transverse noise and reduces the collective coherence, yielding a shallower optimum of $\simeq-8.4$~dB at $t\simeq 31~\mathrm{ms}$. The orange curve shows the corresponding efficiency proxy $\eta_{\mathrm{tot}}(t)$ in the lossy case, which decays steadily during storage (reaching $\eta_{\mathrm{tot}}\simeq0.82$ by $t\simeq125~\mathrm{ms}$), reflecting the progressive reduction of retrievable collective coherence.

\noindent Figure~\ref{fig:fig2}(b) summarizes the best (minimum) retrievable optical squeezing and the readout time $t^*$ at which it occurs as the stored pulse strength $\mu_{\rm in}$ is varied. As $\mu_{\rm in}$ increases from $10^2$ to $2\times10^3$ (equivalently $\theta\simeq 0.02\pi$ to $0.09\pi$ on the top axis), the optimal squeezing improves monotonically (reaching $\sim\!-10$~dB) while the optimal readout time shifts earlier, from $\sim\!10^{-1}$~s down to a few$\times10^{-2}$~s. Including the $^{23}$Na loss channels in the simulation (circles) slightly degrades the best achievable squeezing and favors earlier retrieval compared with the lossless reference (squares). The improvement with increasing $\mu_{\rm in}$ is expected because a larger $\mu_{\rm in}$ prepares a larger initial tilt $\theta$ of the coherent spin state, i.e. a state further from a $J_z$ eigenstate; this increases the available transverse spin length and the spread in $J_z$ that the $\chi J_z^2$ nonlinearity can shear, so stronger squeezing builds up faster before loss and decoherence take over.

Figure~\ref{fig:LO_scan} shows the predicted outcome statistics of a balanced-homodyne readout of the retrieved optical mode as the local-oscillator phase $\phi_{\mathrm{LO}}$ is scanned.  For each $\phi_{\mathrm{LO}}$ we generate $N_{\mathrm{sample}}=10^{6}$ single-shot quadrature outcomes $\epsilon$ by modelling the retrieved field as a displaced Gaussian state and sampling
\begin{equation}
\epsilon=\mu(\phi_{\mathrm{LO}})+\sqrt{V_{\mathrm{det}}(\phi_{\mathrm{LO}})}\,\xi,
\qquad
\xi\sim\mathcal{N}(0,1),
\end{equation}
where the coherent fringe is $\mu(\phi_{\mathrm{LO}})=A_{\mathrm{coh}}\cos(\phi_{\mathrm{LO}}-\phi_{\mathrm{coh}})$ with $A_{\mathrm{coh}}=2\sqrt{\eta_{\mathrm{read}}\,\eta_{\mathrm{coh}}(t)\,\mu_{\mathrm{stored}}}$ and $\mu_{\mathrm{stored}}=\eta_{\mathrm{write}}\zeta_{\mathrm{spatial}}\mu_{\mathrm{in}}$ (clipped to $\le N_0$). Thus the vertical excursion reflects the coherent displacement, while the band thickness at fixed $\phi_{\mathrm{LO}}$ directly encodes the quadrature noise.
The phase-dependent variance is obtained from the transverse collective-spin covariance ellipse at the chosen storage time through Eq.~\ref{eq:Vopt_singlecol}. We overlay the lossless reference (red) and the lossy evolution (blue), evaluating each at the storage time where that case achieves its best squeezing. For visual clarity, the lossy (blue) distribution is displayed with a horizontal offset $\phi_{\mathrm{LO}}\mapsto\phi_{\mathrm{LO}}+\pi/2$ so that the two bands do not obscure one another. In each case the narrowest portion of the band identifies the squeezed quadrature, and its angular offset from the extrema of the coherent fringe reflects the rotation of the noise ellipse familiar from one-axis twisting and optical homodyne detection~\cite{Kitagawa1993,Hosten2016,Lvovsky2009}. Comparing the two overlays highlights how inelastic processes both suppress the coherent modulation (via the reduced effective retrieval amplitude) and broaden/reshape the quadrature-noise band through the modified covariance ellipse.

\section{Conclusions}
We have developed a quantitative theory for generating squeezed optical states using a Bose--Einstein-condensate EIT memory as an interaction-enabled nonlinear element during the storage interval. A weak coherent probe pulse is mapped onto a single, phase-matched collective spin-wave mode of a two-component condensate, preparing a coherent spin state whose initial Bloch-sphere tilt is set by the stored mean excitation number. Collisional interactions then implement one-axis twisting during the hold time, generating transverse spin squeezing that can be exported back into a single, well-defined propagating optical mode on readout.

A key outcome is that the squeezing generation is intrinsically memory-synchronized: the EIT interface fixes the retrieved spatiotemporal mode, while the storage time controls the accumulated nonlinear evolution, enabling on-demand emission of a squeezed optical pulse. Our loss-aware simulations identify an optimal readout time $t^*$ set by the competition between interaction-driven correlations and dissipative processes. In particular, inelastic two-body loss both reduces retrievable coherence and injects transverse noise, shifting the best attainable squeezing to earlier times than in the lossless case; nevertheless, several dB of optical squeezing remain accessible in realistic parameter regimes. This positions the protocol as a hybrid route to mode-matched squeezed light compatible with quantum networking and precision measurement, where controlled release and mode selectivity are often as important as the squeezing level itself.

More broadly, our results connect early proposals that stopped light in condensates could enable nonlinear and quantum optical effects~\cite{Dutton2004, Ginsberg2007} with the capabilities of modern high-efficiency, long-lived BEC memories. Looking ahead, the approach should benefit directly from schemes that extend coherence times and improve the ratio of coherent interaction strength to inelastic loss, for example by optimizing internal-state choices and densities at fixed optical depth, or by leveraging extended-storage platforms such as microgravity BEC memories~\cite{DaRos2023}. In this way, EIT memories offer a unified platform for programmable squeezed-state generation and retrieval, with potential impact for quantum-enhanced optical interferometry and hybrid atom--light metrology.

\section*{Acknowledgements}
M.G. and M.K. acknowledge the support from DLR through funds provided by BMWi (OPTIMO-III, No.~50WM2347), M.G. further acknowledges Einstein Foundation Berlin for support through an Independent Researcher Grant. D.R. acknowledges support by the Federal Ministry of Education and Research of Germany in the project “Open6GHub” (grant number: 16KISK016) and the Deutsche Forschungsgemeinschaft (DFG, German Research Foundation) under Germany’s Excellence Strategy – EXC-2123 QuantumFrontiers – 390837967. This work was supported by the EPSRC International Network in Space Quantum Technologies INSQT (grant ref: EP/W027011/1).

\appendix

\section{Heisenberg correlators used in the OAT simulation}
\label{app:correlators}

We work with two bosonic internal modes $\hat b_{1,2}$ and number operators
$\hat n_i=\hat b_i^\dagger \hat b_i$. The collective spin operators are
\begin{equation}
\hat J_+ = \hat b_2^\dagger \hat b_1,\qquad
\hat J_- = \hat b_1^\dagger \hat b_2,\qquad
\hat J_z = \tfrac12(\hat n_2-\hat n_1).
\label{eq:schwinger_appendix}
\end{equation}

In the Heisenberg picture, all operators are time dependent; for readability we
often omit explicit “$(t)$” on $\hat b_i(t)$ inside the expectation values below.

\subsection{Operator definitions (with bosonic / number-operator forms)}
\label{app:correlators_defs}

The twelve basic correlators are defined as
\begin{align}
E_1(t) &\equiv \langle \hat J_+(t)\rangle
      = \big\langle \hat b_2^\dagger \hat b_1 \big\rangle,
\label{eq:E1_def}\\
E_2(t) &\equiv \langle \hat J_-(t)\rangle
      = \big\langle \hat b_1^\dagger \hat b_2 \big\rangle,
\label{eq:E2_def}\\
E_3(t) &\equiv \langle \hat J_+^2(t)\rangle
      = \big\langle (\hat b_2^\dagger)^2 \hat b_1^2 \big\rangle,
\label{eq:E3_def}\\
E_4(t) &\equiv \langle \hat J_-^2(t)\rangle
      = \big\langle (\hat b_1^\dagger)^2 \hat b_2^2 \big\rangle,
\label{eq:E4_def}
\end{align}
and the number-weighted correlators
\begin{align}
E_5(t) &\equiv \langle \hat n_1\,\hat J_+(t)\rangle
= \big\langle (\hat b_1^\dagger \hat b_1)(\hat b_2^\dagger \hat b_1)\big\rangle,
\label{eq:E5_def}\\
E_6(t) &\equiv \langle \hat n_1\,\hat J_-(t)\rangle
= \big\langle (\hat b_1^\dagger \hat b_1)(\hat b_1^\dagger \hat b_2)\big\rangle,
\label{eq:E6_def}\\
E_7(t) &\equiv \langle \hat n_2\,\hat J_+(t)\rangle
= \big\langle (\hat b_2^\dagger \hat b_2)(\hat b_2^\dagger \hat b_1)\big\rangle,
\label{eq:E7_def}\\
E_8(t) &\equiv \langle \hat n_2\,\hat J_-(t)\rangle
= \big\langle (\hat b_2^\dagger \hat b_2)(\hat b_1^\dagger \hat b_2)\big\rangle,
\label{eq:E8_def}\\
E_9(t) &\equiv \langle \hat J_+(t)\,\hat n_1\rangle
= \big\langle (\hat b_2^\dagger \hat b_1)(\hat b_1^\dagger \hat b_1)\big\rangle,
\label{eq:E9_def}\\
E_{10}(t) &\equiv \langle \hat J_+(t)\,\hat n_2\rangle
= \big\langle (\hat b_2^\dagger \hat b_1)(\hat b_2^\dagger \hat b_2)\big\rangle,
\label{eq:E10_def}\\
E_{11}(t) &\equiv \langle \hat J_-(t)\,\hat n_1\rangle
= \big\langle (\hat b_1^\dagger \hat b_2)(\hat b_1^\dagger \hat b_1)\big\rangle,
\label{eq:E11_def}\\
E_{12}(t) &\equiv \langle \hat J_-(t)\,\hat n_2\rangle
= \big\langle (\hat b_1^\dagger \hat b_2)(\hat b_2^\dagger \hat b_2)\big\rangle.
\label{eq:E12_def}
\end{align}
For $\hat H_{\mathrm{OAT}}=\hbar\chi \hat J_z^2$ the relevant matrix elements acquire
simple $m$-dependent phases. Let the initial state be
$\ket{\theta,\varphi_0}=\sum_{m=-J}^{J}c_m\ket{J,m}$ with
\begin{equation}
c_m
= \sqrt{\binom{2J}{J+m}}\,
\big(\cos\tfrac{\theta}{2}\big)^{J-m}
\big(\sin\tfrac{\theta}{2}\big)^{J+m}
e^{-im\varphi_0},
\label{eq:cm_appendix}
\end{equation}

Define $L_\pm(m)=\sqrt{J(J{+}1)-m(m\mp 1)}$ and $\tau=\chi t$. Then
\begin{align}
E_1(t)
&= \sum_m c_m^*c_{m-1}\,L_+(m)\,e^{+i(2m-1)\tau},
\label{eq:E1_sum}\\
E_2(t)
&= \sum_m c_m^*c_{m+1}\,L_-(m)\,e^{-i(2m+1)\tau},
\label{eq:E2_sum}\\
E_3(t)
&= \sum_m c_m^*c_{m-2}\,L_+(m)L_+(m-1)\,e^{+i(4m-4)\tau},
\label{eq:E3_sum}\\
E_4(t)
&= \sum_m c_m^*c_{m+2}\,L_-(m)L_-(m+1)\,e^{-i(4m+4)\tau},
\label{eq:E4_sum}
\end{align}
and the number-weighted correlators become
\begin{align}
E_5(t)
&= \sum_m (J{-}m)\,c_m^*c_{m-1}\,L_+(m)\,e^{+i(2m-1)\tau},
\\
E_6(t)
&= \sum_m (J{-}m)\,c_m^*c_{m+1}\,L_-(m)\,e^{-i(2m+1)\tau},
\\
E_7(t)
&= \sum_m (J{+}m)\,c_m^*c_{m-1}\,L_+(m)\,e^{+i(2m-1)\tau},
\\
E_8(t)
&= \sum_m (J{+}m)\,c_m^*c_{m+1}\,L_-(m)\,e^{-i(2m+1)\tau},
\\
E_9(t)
&= \sum_m (J{-}m{+}1)\,c_m^*c_{m-1}\,L_+(m)\,e^{+i(2m-1)\tau},
\\
E_{10}(t)
&= \sum_m (J{+}m{-}1)\,c_m^*c_{m-1}\,L_+(m)\,e^{+i(2m-1)\tau},
\\
E_{11}(t)
&= \sum_m (J{-}m{-}1)\,c_m^*c_{m+1}\,L_-(m)\,e^{-i(2m+1)\tau},
\\
E_{12}(t)
&= \sum_m (J{+}m{+}1)\,c_m^*c_{m+1}\,L_-(m)\,e^{-i(2m+1)\tau}.
\end{align}

All sums run over those $m$ values for which the shifted coefficients exist
(e.g. $c_{m-2}$ requires $m\ge -J{+}2$, etc.).

\subsection{Loss-induced rescaling of Heisenberg correlators}
\label{app:loss_rescaling}

Let $E_k^{(u)}(t)$ denote the unitary (lossless) OAT correlators on the fixed-$J$ manifold. To incorporate the reduction of coherence caused by population loss while keeping the OAT phases intact, we use a minimal, number-consistent renormalization based on the mean populations $N_{1,2}(t)$ from Eqs.~\eqref{eq:loss_odes_singlecol}. We define the population ratios $f_1(t)=N_1(t)/N_{1,0}$ and $f_2(t)=N_2(t)/N_{2,0}$, and a coherence ratio motivated by $|\langle \hat J_\pm\rangle|\propto \sqrt{N_1N_2}$ for a phase-coherent two-mode state~\cite{Arecchi1972}, namely $R_1(t)=\sqrt{N_1(t)N_2(t)/(N_{1,0}N_{2,0})}$ and $R_2(t)=R_1^2(t)\equiv\eta_{\mathrm{coh}}$. Here $N_{i,0}=N_i(0)$.

We then rescale the correlators according to their operator content: $E_{1,2}(t)=R_1(t)\,E^{(u)}_{1,2}(t)$, $E_{3,4}(t)=R_2(t)\,E^{(u)}_{3,4}(t)$, $E_{5,6,9,11}(t)=f_1(t)\,R_1(t)\,E^{(u)}_{5,6,9,11}(t)$, and $E_{7,8,10,12}(t)=f_2(t)\,R_1(t)\,E^{(u)}_{7,8,10,12}(t)$. This prescription reduces to the unitary limit when losses vanish. It also ensures that correlators with one (two) ladder operators carry one (two) powers of the coherence factor, while explicit number weighting tracks the corresponding population decay.

To evaluate the diffusion strength $s_q(t)$ used in Eq.~\eqref{eq:LCS_noise_singlecol}, we introduce effective one-, two-, and three-body loss rates $\gamma_{1,2,3}(t)$ obtained by evaluating the corresponding mean-field loss terms on the instantaneous mean state. Writing $N(t)=N_1(t)+N_2(t)$ and defining the mode-overlap integrals of the normalized spatial mode $\phi(\mathbf r)$ as $I_2=\int d^3r\,|\phi(\mathbf r)|^4$ and $I_3=\int d^3r\,|\phi(\mathbf r)|^6$, 
and following Eqs. 14 and 20 of \cite{Li2008}, we write $\gamma_1(t)=[K_{1,1}N_1(t)+K_{1,2}N_2(t)]/N(t)$, $\gamma_2(t)=I_2\,[K_{2,11}N_1(t)+K_{2,22}N_2(t)+\tfrac12 K_{2,12}N(t)]$, and $\gamma_3(t)=I_3\,[K_{3,111}N_1^2(t)+K_{3,222}N_2^2(t)+(K_{3,112}+K_{3,122})N_1(t)N_2(t)]$. $I_2$ and $I_3$ are evaluated by the experimental value of Thomas-Fermi diameter, $d_{\mathrm{TF}}$. The one-, two-, and three-body loss coefficients $K$ are used to model the population dynamics; see Eq.~\ref{eq:loss_odes_singlecol}. The weighted combination entering the Li--Castin--Sinatra diffusion update is then $s_q(t)=\gamma_1(t)+2\gamma_2(t)+3\gamma_3(t)$.

For the simulations used in the article, i.e. for $^{23}$Na in the stretched states $|1\rangle\equiv|F{=}1,m_F{=}{-}1\rangle$ and $|2\rangle\equiv|F{=}2,m_F{=}{-}2\rangle$, we take the one-body loss rate $K_{1,1}=K_{1,2}=2.9\times10^{-2}\,\mathrm{s^{-1}}$ and the three-body coefficients
\begin{align}
K_{3,111} &= 1.57\times10^{-42}\,\mathrm{m^6\,s^{-1}},\\
K_{3,222} &= 1.53\times10^{-41}\,\mathrm{m^6\,s^{-1}}\qquad(\text{upper bound}),
\end{align}
from condensate lifetime measurements in a large-volume optical dipole trap \cite{Gorlitz2003NaF2}, and $K_{2,11}=K_{2,22}=0$; and $K_{2,12}=\frac{8\pi\hbar}{m_{\rm Na}}\,\mathrm{Im}\left(a_{12}\right)$, with $\mathrm{Im}\left(a_{12}\right) = -1\times10^{-3}a_0$ at $B\simeq 132.4~\mathrm{G}$~\cite{Zhang2009}. Mixed-channel three-body coefficients are not available; we approximate them by the geometric mean, $K_{3,112}=K_{3,122}=\sqrt{K_{3,111}K_{3,222}}$.

\section{Mapping between optical coherent amplitude $\alpha$ and atomic coherent spin states}
\label{app:alpha_mapping}

This appendix clarifies the relation between the optical input amplitude $\alpha$ (with $|\alpha|^2=\mu_{\rm in}$), the stored spin-wave coherent amplitude $\beta$, and the rotation/ladder parameterizations of atomic (Bloch) coherent spin states. The key point is that in the weak-excitation (Holstein--Primakoff) regime $\mu_{\rm stored}\ll N_0$, the bosonic spin-wave coherent state $\ket{\beta}_S$ and the atomic coherent spin state $\ket{\theta,\varphi_0}$ describe the same physical state, expressed in two equivalent languages.

\subsection{From optical input to stored spin-wave amplitude}

We take the input probe mode to be in a coherent state $\ket{\alpha}_{\rm in}$ with $|\alpha|^2=\mu_{\rm in}$. For a phase-preserving linear write process with overall mode-matched storage factor $\eta_{\rm write}\zeta_{\rm spatial}$, coherent amplitudes scale with the square root of the efficiency, giving $\beta=\sqrt{\eta_{\rm write}\zeta_{\rm spatial}}\,\alpha$. Consequently, the mean stored excitation number is $\mu_{\rm stored}=|\beta|^2=\eta_{\rm write}\zeta_{\rm spatial}\mu_{\rm in}$, and the stored azimuthal phase is $\varphi_0=\arg\beta$.

\subsection{Spin-wave coherent state and Holstein--Primakoff reduction}

Define the spin-wave vacuum as $\ket{0}_S\equiv\ket{N_0,0}$ and the spin-wave coherent state $\ket{\beta}_S=\hat D_S(\beta)\ket{0}_S=e^{-|\beta|^2/2}\exp(\beta \hat S^\dagger)\ket{0}_S$. In the Holstein--Primakoff regime $\mu_{\rm stored}=\langle \hat S^\dagger\hat S\rangle\ll N_0$, the collective excitation operator may be embedded as $\hat S^\dagger\simeq(\hat b_2^\dagger\hat b_1)/\sqrt{N_0}$. Substituting yields the ladder representation on the fixed-$N_0$ manifold,
\begin{equation}
\ket{\beta}_S \simeq e^{-|\beta|^2/2}\exp\!\Big[\frac{\beta}{\sqrt{N_0}}\,\hat b_2^\dagger\hat b_1\Big]\ket{N_0,0},
\label{eq:beta_HP_ladder_app_short}
\end{equation}
which is the form used in the main text after inserting $\beta=\sqrt{\eta_{\rm write}\zeta_{\rm spatial}}\,\alpha$.

\subsection{Atomic coherent spin states: rotation and ladder forms}

Introduce collective spin operators in the Schwinger representation $\hat J_+=\hat b_2^\dagger\hat b_1$, $\hat J_-=\hat b_1^\dagger\hat b_2$, and $\hat J_z=(\hat b_2^\dagger\hat b_2-\hat b_1^\dagger\hat b_1)/2$, with total spin $J=N_0/2$ and $\ket{N_0,0}\equiv\ket{J,-J}$. 
Atomic coherent spin states are defined as rotations of an extremal Dicke state,
\begin{equation}
\ket{\theta,\varphi}\equiv e^{-i\varphi\hat J_z}e^{-i\theta\hat J_y}\ket{J,-J}.
\label{eq:ACS_rotation_app_short}
\end{equation}
Using the standard disentangling identity for SU(2) rotations, the same state can be written in ladder form as $\ket{\theta,\varphi}=(1+|\tau|^2)^{-J}\exp(\tau\hat J_+)\ket{J,-J}$, where $\tau=e^{i\varphi}\tan(\theta/2)$. Comparing with Eq.~\eqref{eq:beta_HP_ladder_app_short} identifies $\tau=\beta/\sqrt{N_0}$, i.e. $\tau=\sqrt{\eta_{\rm write}\zeta_{\rm spatial}/N_0}\,\alpha$, and $\varphi_0=\arg\beta$.

\subsection{Normalization matching and small-angle limit}

The ladder form includes the CSS normalization $(1+|\tau|^2)^{-J}$. With $J=N_0/2$ and $|\tau|^2=|\beta|^2/N_0=\mu_{\rm stored}/N_0\ll 1$, one has $(1+|\tau|^2)^{-J}\simeq\exp(-J|\tau|^2)=\exp(-|\beta|^2/2)$, which coincides with the bosonic coherent-state prefactor in $\ket{\beta}_S$. Finally, for an atomic coherent state the mean population in state $\ket{2}$ is $\mu_{\rm stored}=\langle \hat n_2\rangle=N_0\sin^2(\theta/2)$, so in the weak-excitation regime $\mu_{\rm stored}\ll N_0$ one finds $\theta\simeq 2\sqrt{\mu_{\rm stored}/N_0}=2\sqrt{\eta_{\rm write}\zeta_{\rm spatial}\mu_{\rm in}/N_0}$.

\section{Suppression of spin demixing in the weak-excitation regime}
\label{app:demixing}

Here we justify a posteriori why spatial phase separation (spin demixing) does not invalidate the collective-spin, single-mode simulations reported in the main text, despite the fact that the sodium interaction parameters adopted from Ref.~\cite{Zhang2009} satisfy the immiscibility condition $a_{12}^2>a_{11}a_{22}$. The key point is that in the weak-excitation (Holstein--Primakoff) regime $N_2\ll N_1$ the demixing instability rate scales linearly with the minority density $n_2$, and the most unstable demixing wavelength grows as $n_2^{-1/2}$, so demixing becomes parametrically slow and can be further suppressed by finite system size.

We summarize the standard linear-stability (Bogoliubov) theory for a homogeneous two-component condensate with equal atomic mass $m$ and repulsive contact interactions $g_{ij}=4\pi\hbar^2 a_{ij}/m$. Linearizing the coupled Gross--Pitaevskii equations around uniform background densities $n_1$ and $n_2$ yields two excitation branches $\Omega_{\pm}(k)$ with~\cite{Kasamatsu2004MI,Timmermans1998PhaseSeparation}
\begin{equation}
\begin{aligned}
\Omega_{\pm}^2(k)
= \varepsilon_k \Bigl[
&\varepsilon_k + g_{11}n_1 + g_{22}n_2 \\
&\pm \sqrt{\left(g_{11}n_1-g_{22}n_2\right)^2
      + 4g_{12}^2 n_1 n_2}
\Bigr],
\end{aligned}
\end{equation}
with $\varepsilon_k = \frac{\hbar^2k^2}{2m}$. For repulsive interactions the long-wavelength miscibility criterion is $g_{12}^2<g_{11}g_{22}$ (equivalently $a_{12}^2<a_{11}a_{22}$), whereas in the immiscible regime $g_{12}^2>g_{11}g_{22}$ the lower branch $\Omega_-(k)$ becomes imaginary for sufficiently small $k$, implying exponential growth of spin-density modulations (modulational instability)~\cite{Kasamatsu2004MI,Timmermans1998PhaseSeparation}. For the sodium parameters quoted by Ref.~\cite{Zhang2009}, $a_{11}=2.8~\mathrm{nm}$, $a_{22}=3.4~\mathrm{nm}$, $a_{12}=3.4~\mathrm{nm}$, one finds $a_{12}^2/(a_{11}a_{22})\simeq 3.4/2.8\simeq 1.21>1$, consistent with the phase-separating behavior reported in Ref.~\cite{Zhang2009} for strong stored pulses.

To extract a demixing timescale, define $A\equiv g_{11}n_1+g_{22}n_2$ and $D\equiv \sqrt{(g_{11}n_1-g_{22}n_2)^2+4g_{12}^2n_1n_2}$, and introduce $\bar\Delta\equiv D-A$. In the immiscible regime one has $\bar\Delta>0$, and the unstable band corresponds to $0<\varepsilon_k<\bar\Delta$, for which $\Omega_-(k)=i\Gamma(k)$ with growth rate $\Gamma(k)=\hbar^{-1}\sqrt{\varepsilon_k(\bar\Delta-\varepsilon_k)}$~\cite{Kasamatsu2004MI}. The maximum growth rate occurs at $\varepsilon_k=\bar\Delta/2$, yielding $\Gamma_{\max}=\bar\Delta/(2\hbar)$ and an $e$-folding time $\tau_{\mathrm{MI}}\equiv \Gamma_{\max}^{-1}=2\hbar/\bar\Delta$. The corresponding most unstable wave number satisfies $\varepsilon_{k_*}=\bar\Delta/2$, i.e.\ $k_*=\sqrt{m\bar\Delta}/\hbar$ and $\lambda_*=2\pi/k_*=2\pi\hbar/\sqrt{m\bar\Delta}$.

In the weak-excitation limit $n_2\ll n_1$ (relevant for $N_2\ll N_1$) one can expand $\bar\Delta$ to first order in $n_2$ and obtain
\begin{equation}
\begin{aligned}
\bar\Delta &\simeq 2n_2\!\left(\frac{g_{12}^2}{g_{11}}-g_{22}\right), \\
\Gamma_{\max} &\simeq \frac{n_2}{\hbar}\!\left(\frac{g_{12}^2}{g_{11}}-g_{22}\right), \\
\tau_{\mathrm{MI}} &\propto \frac{1}{n_2}.
\end{aligned}
\end{equation}
so the demixing instability rate is expected to be parametrically suppressed by the minority density. 

It is worth emphasizing what the weak-excitation limit does (and does not) imply. The immiscibility condition $g_{12}^2>g_{11}g_{22}$ still determines whether the uniform mixture is unstable at long wavelengths. However, in the strongly imbalanced case $n_2\ll n_1$ the dynamically relevant instability is carried by the out-of-phase (spin-density) branch, and its maximum growth rate is proportional to the minority density, $\Gamma_{\max}\propto n_2$ (equivalently $\tau_{\mathrm{MI}}\propto 1/n_2$), while the most unstable wavelength increases as $\lambda_*\propto n_2^{-1/2}$. By contrast, the in-phase density branch remains essentially the majority-component phonon and can scale with $n_1$ without setting the demixing timescale.

In our collective-spin setting the minority population is set by the stored excitation number, with $N_2(0)=N_0\sin^2(\theta/2)\simeq \mu_{\mathrm{eff}}$ for $\mu_{\mathrm{eff}}\ll N_0$, and therefore $n_2\propto N_2\propto \mu_{\mathrm{eff}}$. As a result, $\tau_{\mathrm{MI}}$ grows as $\mu_{\mathrm{eff}}^{-1}$ and the most unstable wavelength grows as $\lambda_*\propto \bar\Delta^{-1/2}\propto n_2^{-1/2}\propto \mu_{\mathrm{eff}}^{-1/2}$, i.e.\ the demixing pattern becomes longer-wavelength and harder to realize as the stored excitation is reduced.

A further stabilizing effect is finite system size. In a trapped cloud of characteristic radius $R$, the smallest accessible wave number is $k_{\min}\sim \pi/R$ and the corresponding kinetic energy is $\varepsilon_{\min}=\hbar^2k_{\min}^2/(2m)$. Since the instability requires $\varepsilon_k<\bar\Delta$, a necessary condition for any unstable mode to fit is $\varepsilon_{\min}<\bar\Delta$, or equivalently $\lambda_*\lesssim 2R$ up to factors of order unity. For the baseline parameters used in our simulations ($N_0=1\times 10^{5}$, $d_{\mathrm{TF}}\simeq 10~\mu\mathrm{m}$ so $R\simeq 5~\mu\mathrm{m}$, and $\mu_{\mathrm{eff}}\sim 10^{3}$), evaluating the above expressions with effective peak densities gives $\Gamma_{\max}^{-1}\sim 15~\mathrm{ms}$ and $\lambda_*\sim 30~\mu\mathrm{m}$, i.e.\ $\lambda_*$ significantly exceeds the cloud diameter and the condition $\varepsilon_{\min}<\bar\Delta$ is not met. Therefore the modulational instability is strongly suppressed on the $\sim 10$--$50~\mathrm{ms}$ timescales relevant to the squeezing dynamics discussed in the main text. For much larger excitations, e.g.\ $\mu_{\mathrm{eff}}\gtrsim 10^{4}$, one expects $\lambda_*$ to decrease toward the system size and demixing to become dynamically relevant, consistent with the strong-pulse regime and larger condensate sizes explored in Ref.~\cite{Zhang2009}.

Finally, our stability estimate based on coupled mean-field Gross--Pitaevskii theory is conservative in that it neglects beyond-mean-field quantum-fluctuation effects, which become especially relevant when a collective (spin-density) mode softens near an instability.  In Bose--Bose mixtures the leading Lee--Huang--Yang correction provides an additional positive ``quantum pressure'' term in the energy functional that stiffens long-wavelength density and spin-density modulations; in particular, it can arrest mean-field instabilities and favor a stable overlapped state with a well-defined composition.  This mechanism underlies the formation of ultradilute quantum droplets predicted in Ref.~\cite{Petrov2015} and observed experimentally in the $^{39}$K mixture~\cite{Cabrera2018}.  While our squeezing simulations do not rely on droplet formation, the same fluctuation-induced stiffness would further reduce the propensity for spatial spin demixing on the millisecond timescales of interest here, complementing the suppression already implied by the small minority density and finite system size discussed above.

\bibliographystyle{apsrev4-2}
\bibliography{main}

@article{Gorshkov2007,
  title = {Photon storage in $\ensuremath{\Lambda}$-type optically dense atomic media. II. Free-space model},
  author = {Gorshkov, Alexey V. and Andr\'e, Axel and Lukin, Mikhail D. and S\o{}rensen, Anders S.},
  journal = {Phys. Rev. A},
  volume = {76},
  issue = {3},
  pages = {033805},
  numpages = {25},
  year = {2007},
  month = {Sep},
  publisher = {American Physical Society},
  doi = {10.1103/PhysRevA.76.033805},
  url = {https://link.aps.org/doi/10.1103/PhysRevA.76.033805}
}

@article{Kuzmich2000,
  title = {Generation of Spin Squeezing via Continuous Quantum Nondemolition Measurement},
  author = {Kuzmich, A. and Mandel, L. and Bigelow, N. P.},
  journal = {Phys. Rev. Lett.},
  volume = {85},
  issue = {8},
  pages = {1594--1597},
  numpages = {0},
  year = {2000},
  month = {Aug},
  publisher = {American Physical Society},
  doi = {10.1103/PhysRevLett.85.1594},
  url = {https://link.aps.org/doi/10.1103/PhysRevLett.85.1594}
}

@article{Saffman2009,
  title = {Spin squeezing of atomic ensembles by multicolor quantum nondemolition measurements},
  author = {Saffman, M. and Oblak, D. and Appel, J. and Polzik, E. S.},
  journal = {Phys. Rev. A},
  volume = {79},
  issue = {2},
  pages = {023831},
  numpages = {8},
  year = {2009},
  month = {Feb},
  publisher = {American Physical Society},
  doi = {10.1103/PhysRevA.79.023831},
  url = {https://link.aps.org/doi/10.1103/PhysRevA.79.023831}
}

@article{Schleier-Smith2010,
  title = {States of an Ensemble of Two-Level Atoms with Reduced Quantum Uncertainty},
  author = {Schleier-Smith, Monika H. and Leroux, Ian D. and Vuleti\'{c}, Vladan},
  journal = {Phys. Rev. Lett.},
  volume = {104},
  issue = {7},
  pages = {073604},
  numpages = {4},
  year = {2010},
  month = {Feb},
  publisher = {American Physical Society},
  doi ={10.1103/PhysRevLett.104.073604},

}

@book{WallsMilburn,
  title = {Quantum Optics},
  author = {Walls, D. F. and Milburn, G. J.},
  publisher = {Springer},
  edition = {2},
  year = {2008}
}

@article{Arecchi1972,
  title = {Atomic coherent states in quantum optics},
  author = {Arecchi, F. T. and Courtens, E. and Gilmore, R. and Thomas, H.},
  journal = {Phys. Rev. A},
  volume = {6},
  pages = {2211--2237},
  year = {1972},
  doi = {10.1103/PhysRevA.6.2211}
}

@article{Fleischhauer2005,
  title = {Electromagnetically induced transparency: Optics in coherent media},
  author = {Fleischhauer, Michael and Imamoglu, Atac and Marangos, Jonathan P.},
  journal = {Rev. Mod. Phys.},
  volume = {77},
  issue = {2},
  pages = {633--673},
  numpages = {0},
  year = {2005},
  month = {Jul},
  publisher = {American Physical Society},
  doi = {10.1103/RevModPhys.77.633},
  url = {https://link.aps.org/doi/10.1103/RevModPhys.77.633}
}

@phdthesis{Choi2011coherent,
  title        = {Coherent control of entanglement with atomic ensembles},
  author       = {Choi, Kyung Soo},
  year         = 2011,
  month        = {May},
  school       = {California Institute of Technology},
  type         = {PhD thesis}
}

@article{Kitagawa1993,
  title = {Squeezed spin states},
  author = {Kitagawa, Masahiro and Ueda, Masahito},
  journal = {Phys. Rev. A},
  volume = {47},
  issue = {6},
  pages = {5138--5143},
  numpages = {0},
  year = {1993},
  month = {Jun},
  publisher = {American Physical Society},
  doi = {10.1103/PhysRevA.47.5138},
  url = {https://link.aps.org/doi/10.1103/PhysRevA.47.5138}
}

@article{Ma2011,
title = {Quantum spin squeezing},
journal = {Physics Reports},
volume = {509},
number = {2},
pages = {89-165},
year = {2011},
issn = {0370-1573},
doi = {https://doi.org/10.1016/j.physrep.2011.08.003},
author = {Jian Ma and Xiaoguang Wang and C.P. Sun and Franco Nori}
}

@article{Ma2017,
doi = {10.1088/2040-8986/19/4/043001},
year = {2017},
month = {feb},
publisher = {IOP Publishing},
volume = {19},
number = {4},
pages = {043001},
author = {Lijun Ma and Oliver Slattery and Xiao Tang},
title = {Optical quantum memory based on electromagnetically induced transparency},
journal = {Journal of Optics}
}

@Article{Esteve2008,
author={Est{\`e}ve, J.
and Gross, C.
and Weller, A.
and Giovanazzi, S.
and Oberthaler, M. K.},
title={Squeezing and entanglement in a Bose--Einstein condensate},
journal={Nature},
year={2008},
month={Oct},
day={01},
volume={455},
number={7217},
pages={1216-1219},
issn={1476-4687},
doi={10.1038/nature07332},
url={https://doi.org/10.1038/nature07332}
}

@Article{Kong2020,
author={Kong, Jia
and Jim{\'e}nez-Mart{\'i}nez, Ricardo
and Troullinou, Charikleia
and Lucivero, Vito Giovanni
and T{\'o}th, G{\'e}za
and Mitchell, Morgan W.},
title={Measurement-induced, spatially-extended entanglement in a hot, strongly-interacting atomic system},
journal={Nature Communications},
year={2020},
month={May},
day={15},
volume={11},
number={1},
pages={2415},
issn={2041-1723},
doi={10.1038/s41467-020-15899-1},
url={https://doi.org/10.1038/s41467-020-15899-1}
}

@Article{Northup2014,
author={Northup, T. E.
and Blatt, R.},
title={Quantum information transfer using photons},
journal={Nature Photonics},
year={2014},
month={May},
day={01},
volume={8},
number={5},
pages={356-363},
issn={1749-4893},
doi={10.1038/nphoton.2014.53},
url={https://doi.org/10.1038/nphoton.2014.53}
}

@article{Fleischhauer2000,
  title = {Dark-State Polaritons in Electromagnetically Induced Transparency},
  author = {Fleischhauer, M. and Lukin, M. D.},
  journal = {Phys. Rev. Lett.},
  volume = {84},
  issue = {22},
  pages = {5094--5097},
  numpages = {0},
  year = {2000},
  month = {May},
  publisher = {American Physical Society},
  doi = {10.1103/PhysRevLett.84.5094},
  url = {https://link.aps.org/doi/10.1103/PhysRevLett.84.5094}
}

@Article{Ginsberg2007,
author={Ginsberg, Naomi S.
and Garner, Sean R.
and Hau, Lene Vestergaard},
title={Coherent control of optical information with matter wave dynamics},
journal={Nature},
year={2007},
month={Feb},
day={01},
volume={445},
number={7128},
pages={623-626},
issn={1476-4687},
doi={10.1038/nature05493},
url={https://doi.org/10.1038/nature05493}
}

@article{Lvovsky2009,
  title = {Continuous-variable optical quantum-state tomography},
  author = {Lvovsky, A. I. and Raymer, M. G.},
  journal = {Rev. Mod. Phys.},
  volume = {81},
  issue = {1},
  pages = {299--332},
  numpages = {0},
  year = {2009},
  month = {Mar},
  publisher = {American Physical Society},
  doi = {10.1103/RevModPhys.81.299},
  url = {https://link.aps.org/doi/10.1103/RevModPhys.81.299}
}

@article{Lettner2011,
  title = {Remote Entanglement between a Single Atom and a Bose-Einstein Condensate},
  author = {Lettner, M. and M\"ucke, M. and Riedl, S. and Vo, C. and Hahn, C. and Baur, S. and Bochmann, J. and Ritter, S. and D\"urr, S. and Rempe, G.},
  journal = {Phys. Rev. Lett.},
  volume = {106},
  issue = {21},
  pages = {210503},
  numpages = {4},
  year = {2011},
  month = {May},
  publisher = {American Physical Society},
  doi = {10.1103/PhysRevLett.106.210503},
  url = {https://link.aps.org/doi/10.1103/PhysRevLett.106.210503}
}

@article{Zhang1990,
  title = {Coherent states: Theory and some applications},
  author = {Zhang, Wei-Min and Feng, Da Hsuan and Gilmore, Robert},
  journal = {Rev. Mod. Phys.},
  volume = {62},
  issue = {4},
  pages = {867--927},
  numpages = {0},
  year = {1990},
  month = {Oct},
  publisher = {American Physical Society},
  doi = {10.1103/RevModPhys.62.867},
  url = {https://link.aps.org/doi/10.1103/RevModPhys.62.867}
}

@article{Zhang2009,
  title = {Creation of Long-Term Coherent Optical Memory via Controlled Nonlinear Interactions in Bose-Einstein Condensates},
  author = {Zhang, Rui and Garner, Sean R. and Hau, Lene Vestergaard},
  journal = {Phys. Rev. Lett.},
  volume = {103},
  issue = {23},
  pages = {233602},
  numpages = {4},
  year = {2009},
  month = {Dec},
  publisher = {American Physical Society},
  doi = {10.1103/PhysRevLett.103.233602},
  url = {https://link.aps.org/doi/10.1103/PhysRevLett.103.233602}
}

@article{Petrov2015,
  title = {Quantum Mechanical Stabilization of a Collapsing Bose-Bose Mixture},
  author = {Petrov, D. S.},
  journal = {Phys. Rev. Lett.},
  volume = {115},
  issue = {15},
  pages = {155302},
  numpages = {5},
  year = {2015},
  month = {Oct},
  publisher = {American Physical Society},
  doi = {10.1103/PhysRevLett.115.155302},
  url = {https://link.aps.org/doi/10.1103/PhysRevLett.115.155302}
}

@article{Cassens2025,
  title = {Entanglement-Enhanced Atomic Gravimeter},
  author = {Cassens, Christophe and Meyer-Hoppe, Bernd and Rasel, Ernst and Klempt, Carsten},
  journal = {Phys. Rev. X},
  volume = {15},
  issue = {1},
  pages = {011029},
  numpages = {7},
  year = {2025},
  month = {Feb},
  publisher = {American Physical Society},
  doi = {10.1103/PhysRevX.15.011029},
  url = {https://link.aps.org/doi/10.1103/PhysRevX.15.011029}
}

@article{Cabrera2018,
author = {C. R. Cabrera  and L. Tanzi  and J. Sanz  and B. Naylor  and P. Thomas  and P. Cheiney  and L. Tarruell },
title = {Quantum liquid droplets in a mixture of Bose-Einstein condensates},
journal = {Science},
volume = {359},
number = {6373},
pages = {301-304},
year = {2018},
doi = {10.1126/science.aao5686},
URL = {https://www.science.org/doi/abs/10.1126/science.aao5686}}

@article{Kasamatsu2004MI,
  title = {Multiple Domain Formation Induced by Modulation Instability in Two-Component Bose-Einstein Condensates},
  author = {Kasamatsu, Kenichi and Tsubota, Makoto},
  journal = {Phys. Rev. Lett.},
  volume = {93},
  issue = {10},
  pages = {100402},
  numpages = {4},
  year = {2004},
  month = {Sep},
  publisher = {American Physical Society},
  doi = {10.1103/PhysRevLett.93.100402},
}

@article{Timmermans1998PhaseSeparation,
  title   = {Phase Separation of Bose-Einstein Condensates},
  author  = {Timmermans, E.},
  journal = {Phys. Rev. Lett.},
  volume  = {81},
  pages   = {5718--5721},
  year    = {1998},
  doi     = {10.1103/PhysRevLett.81.5718},
}

@article{DaRos2023,
  title = {Proposal for a long-lived quantum memory using matter-wave optics with Bose-Einstein condensates in microgravity},
  author = {Da Ros, Elisa and Kanthak, Simon and Sa\ifmmode \breve{g}\else \u{g}\fi{}lamy\"urek, Erhan and G\"undo\ifmmode \breve{g}\else \u{g}\fi{}an, Mustafa and Krutzik, Markus},
  journal = {Phys. Rev. Res.},
  volume = {5},
  issue = {3},
  pages = {033003},
  numpages = {10},
  year = {2023},
  month = {Jul},
  publisher = {American Physical Society},
  doi = {10.1103/PhysRevResearch.5.033003},
  url = {https://link.aps.org/doi/10.1103/PhysRevResearch.5.033003}
}

@article{Barzel2024,
  doi = {10.22331/q-2024-02-29-1273},
  url = {https://doi.org/10.22331/q-2024-02-29-1273},
  title = {Entanglement dynamics of photon pairs and quantum memories in the gravitational field of the earth},
  author = {Barzel, Roy and G{\"{u}}ndo{\u{g}}an, Mustafa and Krutzik, Markus and R{\"{a}}tzel, Dennis and L{\"{a}}mmerzahl, Claus},
  journal = {{Quantum}},
  issn = {2521-327X},
  publisher = {{Verein zur F{\"{o}}rderung des Open Access Publizierens in den Quantenwissenschaften}},
  volume = {8},
  pages = {1273},
  month = feb,
  year = {2024}
}

@Article{Choi2008,
author={Choi, K. S.
and Deng, H.
and Laurat, J.
and Kimble, H. J.},
title={Mapping photonic entanglement into and out of a quantum memory},
journal={Nature},
year={2008},
month={Mar},
day={01},
volume={452},
number={7183},
pages={67-71},
issn={1476-4687},
doi={10.1038/nature06670},
url={https://doi.org/10.1038/nature06670}
}

@Article{Katz2018,
author={Katz, Or
and Firstenberg, Ofer},
title={Light storage for one second in room-temperature alkali vapor},
journal={Nat. Commun.},
year={2018},
volume={9},
number={1},
pages={2074},
issn={2041-1723},
doi={10.1038/s41467-018-04458-4},
url={https://doi.org/10.1038/s41467-018-04458-4}
}

@article{Seri2017,
  title = {Quantum Correlations between Single Telecom Photons and a Multimode On-Demand Solid-State Quantum Memory},
  author = {Seri, Alessandro and Lenhard, Andreas and Riel\"ander, Daniel and G\"undo\ifmmode \breve{g}\else \u{g}\fi{}an, Mustafa and Ledingham, Patrick M. and Mazzera, Margherita and de Riedmatten, Hugues},
  journal = {Phys. Rev. X},
  volume = {7},
  issue = {2},
  pages = {021028},
  numpages = {7},
  year = {2017},
  month = {May},
  publisher = {American Physical Society},
  doi = {10.1103/PhysRevX.7.021028},
  url = {https://link.aps.org/doi/10.1103/PhysRevX.7.021028}
}

@article{Lei2023,
author = {Yisheng Lei and Faezeh Kimiaee Asadi and Tian Zhong and Alex Kuzmich and Christoph Simon and Mahdi Hosseini},
journal = {Optica},
number = {11},
pages = {1511--1528},
publisher = {Optica Publishing Group},
title = {Quantum optical memory for entanglement distribution},
volume = {10},
month = {Nov},
year = {2023},
url = {https://opg.optica.org/optica/abstract.cfm?URI=optica-10-11-1511},
doi = {10.1364/OPTICA.493732},
}

@article{Harris1990,
  title = {Nonlinear optical processes using electromagnetically induced transparency},
  author = {Harris, S. E. and Field, J. E. and Imamo\u{g}lu, A.},
  journal = {Phys. Rev. Lett.},
  volume = {64},
  issue = {10},
  pages = {1107--1110},
  numpages = {0},
  year = {1990},
  month = {Mar},
  publisher = {American Physical Society},
  doi = {10.1103/PhysRevLett.64.1107},
  url = {https://link.aps.org/doi/10.1103/PhysRevLett.64.1107}
}

@article{Saglamyurek2021,
doi = {10.1088/1367-2630/abf1d9},
url = {https://dx.doi.org/10.1088/1367-2630/abf1d9},
year = {2021},
month = {apr},
publisher = {IOP Publishing},
volume = {23},
number = {4},
pages = {043028},
author = {Erhan Saglamyurek and Taras Hrushevskyi and Anindya Rastogi and Logan W Cooke and Benjamin D Smith and Lindsay J LeBlanc},
title = {Storing short single-photon-level optical pulses in Bose–Einstein condensates for high-performance quantum memory},
journal = {New Journal of Physics}
}

@article{Dutton2004,
  title = {Storing and processing optical information with ultraslow light in Bose-Einstein condensates},
  author = {Dutton, Zachary and Hau, Lene Vestergaard},
  journal = {Phys. Rev. A},
  volume = {70},
  issue = {5},
  pages = {053831},
  numpages = {19},
  year = {2004},
  month = {Nov},
  publisher = {American Physical Society},
  doi = {10.1103/PhysRevA.70.053831},
  url = {https://link.aps.org/doi/10.1103/PhysRevA.70.053831}
}

@Article{Ma2021,
author={Ma, Yu
and Ma, You-Zhi
and Zhou, Zong-Quan
and Li, Chuan-Feng
and Guo, Guang-Can},
title={One-hour coherent optical storage in an atomic frequency comb memory},
journal={Nat. Commun.},
year={2021},
month={Apr},
day={22},
volume={12},
number={1},
pages={2381},
issn={2041-1723},
doi={10.1038/s41467-021-22706-y},
url={https://doi.org/10.1038/s41467-021-22706-y}
}

@Article{Sorensen2001,
author={S{\o}rensen, A.
and Duan, L.-M.
and Cirac, J. I.
and Zoller, P.},
title={Many-particle entanglement with Bose--Einstein condensates},
journal={Nature},
year={2001},
month={Jan},
day={01},
volume={409},
number={6816},
pages={63-66},
issn={1476-4687},
doi={10.1038/35051038},
url={https://doi.org/10.1038/35051038}
}

@article{Boller1991,
  title = {Observation of electromagnetically induced transparency},
  author = {Boller, K.-J. and Imamo\u{g}lu, A. and Harris, S. E.},
  journal = {Phys. Rev. Lett.},
  volume = {66},
  issue = {20},
  pages = {2593--2596},
  numpages = {0},
  year = {1991},
  month = {May},
  publisher = {American Physical Society},
  doi = {10.1103/PhysRevLett.66.2593},
  url = {https://link.aps.org/doi/10.1103/PhysRevLett.66.2593}
}

@Article{Hosten2016b,
author={Hosten, Onur
and Engelsen, Nils J.
and Krishnakumar, Rajiv
and Kasevich, Mark A.},
title={Measurement noise 100 times lower than the quantum-projection limit using entangled atoms},
journal={Nature},
year={2016},
month={Jan},
day={01},
volume={529},
number={7587},
pages={505-508},
issn={1476-4687},
doi={10.1038/nature16176},
url={https://doi.org/10.1038/nature16176}
}

@article{Gross2010,
  author  = {Gross, C. and Zibold, T. and Nicklas, E. and Est{\`e}ve, J. and Oberthaler, M. K.},
  title   = {Nonlinear atom interferometer surpasses classical precision limit},
  journal = {Nature},
  volume  = {464},
  pages   = {1165--1169},
  year    = {2010},
  doi     = {10.1038/nature08919}
}

@Article{Riedel2010,
author={Riedel, Max F.
and B{\"o}hi, Pascal
and Li, Yun
and H{\"a}nsch, Theodor W.
and Sinatra, Alice
and Treutlein, Philipp},
title={Atom-chip-based generation of entanglement for quantum metrology},
journal={Nature},
year={2010},
month={Apr},
day={01},
volume={464},
number={7292},
pages={1170-1173},
issn={1476-4687},
doi={10.1038/nature08988},
url={https://doi.org/10.1038/nature08988}
}

@article{Berrada2013,
  author  = {Berrada, T. and van Frank, S. and B{\"u}cker, R. and Schumm, T. and Schaff, J.-F. and Schmiedmayer, J.},
  title   = {Integrated Mach--Zehnder interferometer for Bose--Einstein condensates},
  journal = {Nature Communications},
  volume  = {4},
  pages   = {2077},
  year    = {2013},
  doi     = {10.1038/ncomms3077}
}

@article{Muessel2014,
  author  = {Muessel, W. and Strobel, H. and Linnemann, D. and Hume, D. B. and Oberthaler, M. K.},
  title   = {Scalable Spin Squeezing for Quantum-Enhanced Magnetometry with Bose-Einstein Condensates},
  journal = {Physical Review Letters},
  volume  = {113},
  pages   = {103004},
  year    = {2014},
  doi     = {10.1103/PhysRevLett.113.103004}
}

@article{Laudat2018,
  author  = {Laudat, T. and Dugrain, V. and Mazzoni, T. and Huang, C.-L. and Garrido Alzar, C. L. and Sinatra, A. and Rosenbusch, P. and Reichel, J.},
  title   = {Spontaneous spin squeezing in a rubidium Bose-Einstein condensate},
  journal = {New Journal of Physics},
  volume  = {20},
  pages   = {073018},
  year    = {2018},
  doi     = {10.1088/1367-2630/aacf1e}
}

@article{Lucke2011,
  author  = {L{\"u}cke, B. and Scherer, M. and Kruse, J. and Pezz{\'e}, L. and Deuretzbacher, F. and Hyllus, P. and T{\'o}pic, O. and Peise, J. and Ertmer, W. and Arlt, J. and Santos, L. and Smerzi, A. and Klempt, C.},
  title   = {Twin Matter Waves for Interferometry Beyond the Classical Limit},
  journal = {Science},
  volume  = {334},
  number  = {6057},
  pages   = {773--776},
  year    = {2011},
  doi     = {10.1126/science.1208798}
}

@article{Hamley2012,
  author  = {Hamley, C. D. and Gerving, C. S. and Hoang, T. M. and Bookjans, E. M. and Chapman, M. S.},
  title   = {Spin-nematic squeezed vacuum in a quantum gas},
  journal = {Nature Physics},
  volume  = {8},
  pages   = {305--308},
  year    = {2012},
  doi     = {10.1038/nphys2245}
}

@article{Fadel2018,
  author  = {Fadel, M. and Zibold, T. and D{\'e}camps, B. and Treutlein, P.},
  title   = {Spatial entanglement patterns and Einstein-Podolsky-Rosen steering in a Bose-Einstein condensate},
  journal = {Science},
  volume  = {360},
  pages   = {409--413},
  year    = {2018},
  doi     = {10.1126/science.aao1850}
}

@article{Kunkel2018,
  author  = {Kunkel, P. and Pr{\"u}fer, M. and Strobel, H. and Linnemann, D. and Fr{\"o}lian, A. and Gasenzer, T. and G{\"a}rttner, M. and Oberthaler, M. K.},
  title   = {Spatially distributed multipartite entanglement enables EPR steering of atomic clouds},
  journal = {Science},
  volume  = {360},
  number  = {6387},
  pages   = {413--416},
  year    = {2018},
  doi     = {10.1126/science.aao2254}
}

@Article{Hau1999,
author={Hau, Lene Vestergaard
and Harris, S. E.
and Dutton, Zachary
and Behroozi, Cyrus H.},
title={Light speed reduction to 17 metres per second in an ultracold atomic gas},
journal={Nature},
year={1999},
month={Feb},
day={01},
volume={397},
number={6720},
pages={594-598},
issn={1476-4687},
doi={10.1038/17561},
url={https://doi.org/10.1038/17561}
}

@article{Liu2001,
  author  = {Liu, Chien and Dutton, Zachary and Behroozi, Cyrus H. and Hau, Lene Vestergaard},
  title   = {Observation of coherent optical information storage in an atomic medium using halted light pulses},
  journal = {Nature},
  volume  = {409},
  number  = {6819},
  pages   = {490--493},
  year    = {2001},
  doi     = {10.1038/35054017}
}

@article{Hosten2016,
author = {O. Hosten  and R. Krishnakumar  and N. J. Engelsen  and M. A. Kasevich },
title = {Quantum phase magnification},
journal = {Science},
volume = {352},
number = {6293},
pages = {1552-1555},
year = {2016},
doi = {10.1126/science.aaf3397}}

@article{Gorlitz2003NaF2,
  author       = {G{\"o}rlitz, A. and Gustavson, T. L. and Leanhardt, A. E. and
                  L{\"o}w, R. and Chikkatur, A. P. and Gupta, S. and Inouye, S. and
                  Pritchard, D. E. and Ketterle, W.},
  title        = {Sodium Bose-Einstein Condensates in the {$F=2$} State in a Large-Volume Optical Trap},
  journal      = {Phys. Rev. Lett.},
  volume       = {90},
  pages        = {090401},
  year         = {2003},
  doi          = {10.1103/PhysRevLett.90.090401},
  eprint       = {cond-mat/0208385},
  archivePrefix= {arXiv}
}

@article{Honda2008,
  title = {Storage and Retrieval of a Squeezed Vacuum},
  author = {Honda, Kazuhito and Akamatsu, Daisuke and Arikawa, Manabu and Yokoi, Yoshihiko and Akiba, Keiichirou and Nagatsuka, Satoshi and Tanimura, Takahito and Furusawa, Akira and Kozuma, Mikio},
  journal = {Phys. Rev. Lett.},
  volume = {100},
  issue = {9},
  pages = {093601},
  numpages = {4},
  year = {2008},
  month = {Mar},
  publisher = {American Physical Society},
  doi = {10.1103/PhysRevLett.100.093601},
  url = {https://link.aps.org/doi/10.1103/PhysRevLett.100.093601}
}

@article{Appel2008,
  title = {Quantum Memory for Squeezed Light},
  author = {Appel, J\"urgen and Figueroa, Eden and Korystov, Dmitry and Lobino, M. and Lvovsky, A. I.},
  journal = {Phys. Rev. Lett.},
  volume = {100},
  issue = {9},
  pages = {093602},
  numpages = {4},
  year = {2008},
  month = {Mar},
  publisher = {American Physical Society},
  doi = {10.1103/PhysRevLett.100.093602},
  url = {https://link.aps.org/doi/10.1103/PhysRevLett.100.093602}
}

@article{Schmied2016,
author = {Roman Schmied  and Jean-Daniel Bancal  and Baptiste Allard  and Matteo Fadel  and Valerio Scarani  and Philipp Treutlein  and Nicolas Sangouard },
title = {Bell correlations in a Bose-Einstein condensate},
journal = {Science},
volume = {352},
number = {6284},
pages = {441-444},
year = {2016},
doi = {10.1126/science.aad8665}}

@article{Li2008,
  title = {Optimum Spin Squeezing in Bose-Einstein Condensates with Particle Losses},
  author = {Li, Yun and Castin, Y. and Sinatra, A.},
  journal = {Phys. Rev. Lett.},
  volume = {100},
  issue = {21},
  pages = {210401},
  numpages = {4},
  year = {2008},
  month = {May},
  publisher = {American Physical Society},
  doi = {10.1103/PhysRevLett.100.210401},
  url = {https://link.aps.org/doi/10.1103/PhysRevLett.100.210401}
}

@Article{Li2009,
author={Li, Yun
and Treutlein, P.
and Reichel, J.
and Sinatra, A.},
title={Spin squeezing in a bimodal condensate: spatial dynamics and particle losses},
journal={The European Physical Journal B},
year={2009},
month={Apr},
day={01},
volume={68},
number={3},
pages={365-381},
issn={1434-6036},
doi={10.1140/epjb/e2008-00472-6},
url={https://doi.org/10.1140/epjb/e2008-00472-6}
}

\end{document}